\documentclass{article}
\usepackage{arxiv}
\usepackage{jabbrv}
\usepackage[numbers, sort&compress]{natbib}
\usepackage{color, soul}
\usepackage{zref-xr}
\zxrsetup{toltxlabel=true, tozreflabel=false}
\usepackage[utf8]{inputenc} 
\usepackage[T1]{fontenc}    
\usepackage{url}            
\usepackage{booktabs}       
\usepackage{amsfonts}       
\usepackage{graphicx}
\usepackage{doi}
\usepackage{siunitx}
\usepackage{wrapfig}
\usepackage[hang]{subfigure}
\usepackage{amsmath}
\usepackage{caption}
\usepackage[most]{tcolorbox}
\usepackage[export]{adjustbox}
\usepackage{floatrow}
\usepackage{placeins}

\newenvironment{stretchpars}
 {\par\setlength{\parfillskip}{0pt}}
 {\par}

\title{Quantifying Tipping Risks in Power Grids and beyond}

\usepackage{hyperref}
\zexternaldocument*{SI_Quantifying_Tipping_Risks}

\author{ \href{https://orcid.org/0000-0003-0529-7926}{\includegraphics[scale=0.06]{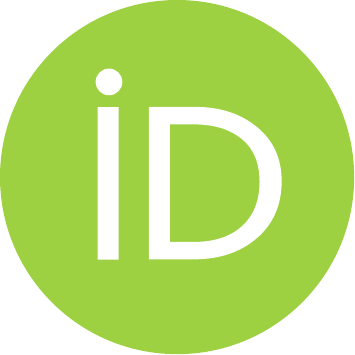}\hspace{1mm}Martin Heßler}\thanks{Center for Nonlinear Science, Westphalian Wilhelms-University Münster, 48149 Münster, Germany} \\
	Institute for Theoretical Physics\\
	Westphalian Wilhelms-University Münster\\
	48149 Münster, North Rhine-Westphalia, Germany\\
	\texttt{m\_{}hess23@uni-muenster.de} \\
	\And
	\href{https://orcid.org/0000-0003-0986-0878 }{\includegraphics[scale=0.06]{orcid.pdf}\hspace{1mm}Oliver Kamps} \\
	Center for Nonlinear Science\\
	Westphalian Wilhelms-University Münster\\ 
	48149 Münster, North Rhine-Westphalia, Germany\\
	\texttt{okamp@uni-muenster.de} \\
}

\renewcommand{\headeright}{\emph{arXiv} Preprint}
\renewcommand{\undertitle}{\emph{arXiv} Preprint}
\renewcommand{\shorttitle}{Quantifying Tipping Risks}

\hypersetup{
pdftitle={Quantifying Tipping Risks in Power Grids and beyond},
pdfsubject={tipping risk, regime shift, early warning signal, leading indicator, resilience, power grid, cascading failure, power outage, North America Western Interconnection, Bayesian inference, complex system},
pdfauthor={Martin Heßler, Oliver Kamps},
pdfkeywords={tipping risk, regime shift, early warning signal, leading indicator, resilience, power grid, cascading failure, power outage, North America Western Interconnection, Bayesian inference, complex system},
}

\begin{document}
\maketitle

\begin{abstract}
Critical transitions, ubiquitous in nature and technology, necessitate anticipation to avert adverse outcomes. While many studies focus on bifurcation-induced tipping, where a control parameter change leads to destabilization, alternative scenarios are conceivable, e.g. noise-induced tipping by an increasing noise level in a multi-stable system. Although the generating mechanisms can be different, the observed time series can exhibit similar characteristics. Therefore, we propose a Bayesian Langevin approach, implemented in an open-source tool, which is capable of quantifying both deterministic and intrinsic stochastic dynamics simultaneously. After a detailed proof of concept, we analyse two bus voltage frequency time series of the historic North America Western Interconnection blackout on 10th August 1996. Our results unveil the intricate interplay of changing resilience and noise influence. A comparison with the blackout’s timeline supports our frequency dynamics’ Langevin model, with the BL-estimation indicating a permanent grid state change already two minutes before the officially defined triggering event. A tree-related high impedance fault or sudden load increases may serve as earlier triggers during this event, as suggested by our findings. This study underscores the importance of distinguishing destabilizing factors for a reliable anticipation of critical transitions, offering a tool for better understanding such events across various disciplines.
\end{abstract}


\section{Introduction}\label{sec: intro}
Complex dynamical systems like Earth's global climate, ecosystems, the human brain, or infrastructure such as power grids and communication systems are composed of a large number of interacting parts which can operate on various scales in space and time. 
Often the observation of such a system gives only partial information about the involved processes and it is not simply possible to identify the reason leading to a certain behaviour.  
In this context very prominent phenomena are sudden transitions, so-called tipping events, where the system undergoes a fast transition into a qualitatively different state. Unfortunately, there are numerous possible pathways to tipping events that can differ strongly in the key mechanisms. This fact complicates the anticipation of such critical transitions significantly which is desired in a wide range of research fields and everyday life to mitigate or prevent damage and in order to control complex systems.~\cite{a:veraart11, a:Corrado2014, a:livina2010,a:Livina2015, a:Lenton2012, a:Rikkert2016, a:Dakos2014, a:izrailtyan, a:Leemput2013, a:Hessler2023b} To achieve these goals research is done to develop stability measures or \textit{leading indicators} calculated on time series data that are applicable in the absence of detailed knowledge about the underlying laws of the system dynamics.~\cite{a:dakos12, a:carpenter11, a:livina2013, a:Liang2017, a:Xie2018, a:Dakos2012}\\
Common leading indicator candidates are designed focused on \textit{critical slowing down} (CSD) or \textit{flickering} that are mentioned as general phenomena only in connection with uprising bifurcation-induced transitions (B-tipping).~\cite{a:scheffer09, a:scheffer12} Briefly summarized, CSD denotes the phenomenon of an increasing relaxation time prior to a bifurcation which can e.g. lead to higher lag-1 autocorrelation (AR1) $\hat{\rho}$ and standard deviation (std) $\tilde{\sigma}$ of a time series. Ongoing jumps between two branches in a bi-stable regime are called flickering and can cause an increasing skewness of the data distribution. However, since the common leading indicators rely on CSD and flickering, they are limited by design to the above mentioned B-tipping scenario which is only one of numerous routes to a tipping event: Additionally to B-tipping, a fast variation of the control parameter can also lead to \textit{rate-dependent tipping} (R-tipping) when the phase space is modified much faster than the time scale on which the system is able to relax onto the modified stable branch (cf. supplementary material~\cite{a:supplNtip}, section \ref{sec:R tipping}). Furthermore, the overall noise level has to be small enough to avoid noise-triggered jumps to alternative stable states which are also known as \textit{noise-induced} tipping~\cite{a:Ritchie2017} events (N-tipping). Note that N-tipping can already take place for parameter configurations that lie far away from a critical B-tipping threshold. Such N-tipping is intrinsically hard to predict, but it is an important candidate of driving tipping events in many natural and technical systems, e.g. power grids.~\cite{a:Ritchie2017, a:cotilla-sanchez12, a:ren2015} In the framework of graph theory and networks~\cite{b:bollobas1998, a:doerfler2018} power grids can be modeled by coupled nonlinear oscillators, such as e.g. provided by the Kuramoto model~\cite{a:Guo2021, a:Filatrella2008, a:Grzybowski2016}. Their operating regimes are threatened by the risk of de-synchronization which leads to power outages. Such outages can have occasionally severe consequences for industry and private consumers. Fortunately, unlike in many other sciences, e.g. ecology~\cite{a:bissonette1999}, highly resolved data of characteristic observables, e.g. bus voltage frequency time series --- as they are needed in general for leading indicator analyses --- are easily available today.\\ 
Previous studies focused on destabilization due to B-tipping.~\cite{a:cotilla-sanchez12, a:ren2015} However, power grids are permanently exposed to intrinsic variable noise, e.g. from renewable energy sources as wind power or solar stations (cf. Infobox 1, figure \ref{fig: control scheme} (a)).\cite{a:Gorjao2021, a:milan2013} Furthermore, the currently very important trend to an environmentally sustainable grid architecture will probably augment the amount of noisy grid participants significantly in future. Under these circumstances, N-tipping, or at least the influence of internal noise, can play an important role in destabilizing power grids. It would be worthwhile to have the opportunity to monitor changes in both resilience and noise levels simultaneously at a given time resolution to effectively control the system, better understand the frequency dynamics before power outage events, and, in rare cases, possibly avoid power outages.
In principle, a model based on a Langevin equation 
\begin{align}
\dot{\underline{x}}(\underline{x},t) = \underline{h}(\underline{x}(t),t) + \underline{g}(\underline{x}(t), t) \underline{\Gamma} (t),\label{eq:langevin}
\end{align}
with the drift $\underline{h}(\underline{x}(t),t)$, the diffusion $\underline{g}(\underline{x}(t), t)$ and stochasticity $\underline{\Gamma } (t)$ is able to do so.
It provides access to B-tipping risk factors which are included in the drift term and N-tipping risk factors which are modelled by the diffusion term for a given time scale resolution. We provide an illustrative introduction to the theoretical ideas behind modelling frequency dynamics via the Langevin approach in Infobox 1, figure \ref{fig: control scheme}. 
Typically, power grids (cf. figure \ref{fig: control scheme} (a)) consist of various structured levels reaching from highest voltage grids, e.g. the transmissive supergrid of a country or continent, to local low voltage grids supplying cities and households with electricity. Additionally to the levels being connected to each other via electric tension transformation, there are different intra-level suppliers on the one hand, e.g. nuclear or fossil power plants and energy trading on the supergrid level, on-shore wind parks or solar parks on the regional scale, and consumers on the other hand, e.g. large-scale industries and railway on the high-voltage grid and so on. The total balance of power input and output is crucial for the power grids' stable synchrony of usually  \SI{60}{\hertz} (e.g. USA, Japan) or \SI{50}{\hertz} (e.g. Europe, again Japan) bus voltage frequency. The grid components act partially on a controllable slow time scale $\tau_{\rm det}$ which can thus be modelled in a deterministic fashion. These components' dynamics are basically modelled by the deterministic drift $\underline{h}(\underline{x}(t),t)$. We approximately assign the components which act (partially) on a controllable time scale $\tau_{\rm det}$ by a gray (gray-orange) tile in Infobox 1, figure \ref{fig: control scheme}. For example this could be the controlled (de-)activation of a power plant or industrial consumer as well as the railway grid under normal conditions. Nevertheless, industrial and scientific stakeholders only report activations or deactivations up to a certain extent, and railway traffic can be heavily disrupted, for example, due to damage to the railway infrastructure caused by a storm or unforeseen strikes. Note that power outages as well as control mechanisms live on time scales of seconds up to several minutes, e.g. global grid primary control actions in Germany react to a maximum frequency deviation of $\pm\SI{0.2}{\hertz}$ in times up to \SI{30}{\second} and have to be stable under full load up to \SI{15}{\minute} before the simultaneously starting secondary control in regional grids should replace the primary control completely.~\cite{a:primarycontrol} For these reasons we neglect for example the daily-periodic solar energy supply or the on-average seasonal wind energy supply, since they live on time scales much less relevant for power outage events even if they could be easily incorporated in a deterministic drift model of daily up to monthly time scales. Accordingly, they should be compensated in the considered bus voltage frequency dynamics. In this spirit, the solar park example is mainly driven by additional phenomena as a variable cloudy and sunny sky and wind parks depend crucial on turbulent fluctuations in the atmosphere which live on faster time scales, assigned by orange tiles.~\cite{a:Gorjao2021}\\
The stable state $\underline{x}^*$ of the normally operating power grid is theoretically reflected by a valley of the potential $\underline{V}(\underline{x}) = - \int \underline{h}(\underline{x}(t),t){\rm d}\underline{x}$ the mountainsides of which determine the strength of the restoring force by their steepness. A schematic illustration is given in Infobox 1, figure \ref{fig: control scheme} (b.1). Each available option of a control action and each infrastructure extension essentially affects the potential landscape. An unforeseen failure of a component can even flip the stable potential valley to an unstable mountain top leading to a B-tipping event (cf. Infobox 1, figure \ref{fig: control scheme} (b.1)) or at least decrease the potential barrier's height making N-tipping more probable (cf. Infobox 2, figure \ref{fig: mixed scenario scheme}). All in all, if the\vspace{-2mm}\begin{stretchpars}dominant processes live on the time scale $\tau_{\rm \det}$ the power grid states' pdf should be rather peaked around the stable\end{stretchpars}
\begin{tcolorbox}[title=Infobox 1: Power Grids{,} Frequency Dynamics and the Langevin Model,
title filled=false,
colback=white!5!white,
colframe=black!75!black,
boxsep = 0.3mm,
fonttitle = \bfseries,
bottom = 3mm]
\begin{wrapfigure}{l}{0.65\textwidth}
\includegraphics[width=\textwidth]{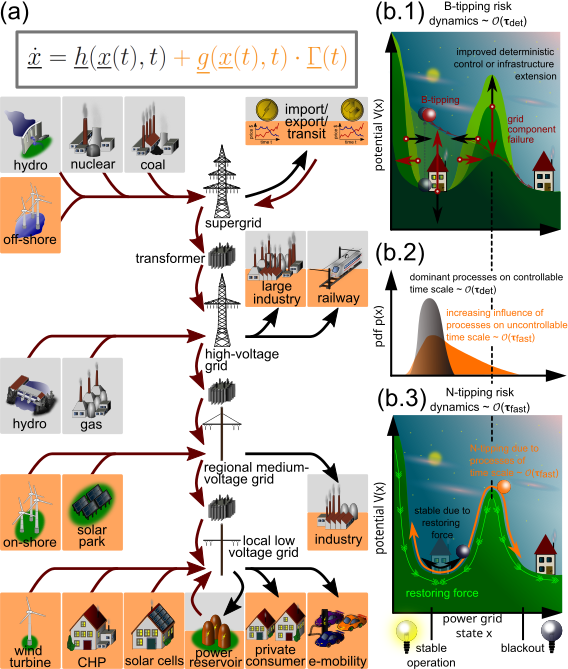}
\refstepcounter{figure}
\label{fig: control scheme}
\end{wrapfigure}
\noindent\normalsize{\bfseries\textsf{Figure} ~\thefigure.} Scheme to illustrate the relation between the bus voltage frequency dynamics in power grids and the Langevin model approach. The most frequent and important components of each grid level are shown for the sake of clarity. In general, overlaps between the levels of grid components are possible. (a) A power grid typically consists of various voltage levels from highest-voltage supergrid to low voltage local grid scales connected by electric tension transformers. The most prominent grid suppliers and consumers which are shown in the scheme, live on various time scales. Their dynamics can only be controlled and modelled by the deterministic drift $\underline{h}(\underline{x}(t), t)$ if they live on a suitable slow time scale $\tau_{\rm det}$. Phenomena of the grid participants on shorter time scales $\tau_{\rm fast}$ have to be captured by the diffusion $\underline{g}(\underline{x}(t),t)\cdot \underline{\Gamma}(t)$ in form of intrinsic stochastic dynamics. We assign the suitable time scales to the components with gray tiles for $\tau_{\rm det}$, orange tiles for $\tau_{\rm fast}$ and mixed tiles for components that partially contribute in each manner. The assignment is only an approximation and serves for illustration purposes. (b.1) From a modeller's point of view the power grid in stable operation is located in the minimum of the potential $\underline{V}(\underline{x}) = - \int \underline{h}(\underline{x}(t),t){\rm d}\underline{x}$. 
Whereas the stable operation state of the power system is a fixed point in this mathematical framework, the projection of the imaginable ensemble of power grid states described on the microscopic topological scale into one dimension is a simplification that serves for illustration purposes. Nevertheless, the macroscopic frequency dynamics is in fact basically one-dimensional around the fixed point of stable operation and aggregates most of the information from lower scales. 
In reality, the power outage landscape on the right side of the mountain is unknown terrain and rather corresponds to a periodic, chaotic attractor or a transient dynamics. The landscape describes the current grid infrastructure under possible control. Metaphorically speaking, the system state can be imagined as a ball driven by stochastic kicks and the deterministic restoring gravitation force in the potential landscape. A component failure can negatively modify the potential landscape by decreasing the potential barrier, flattening or even flipping the stable potential valley into a mountain top or mountainside (shaded potential landscape) which would correspond to a B-tipping event. (b.2) If the power grid's control works properly for the given infrastructure, the probability density function (pdf) $p(x)$ of possible states is peaked around the stable operation state $x^*$ as illustrated by the gray shaded pdf. Increasing influence of uncontrollable stochasticity, for example, because of a higher influence of the orange-tiled fast scale components in (a), can widen the possible range of states as indicated by the orange pdf. (b.3) If the restoring force (light green arrow line) given by the steepness of the potential mountainsides is high enough, the power grid state fluctuates in principle only slightly around the stable operation state (black arrows corresponding to the width of the gray pdf in (b.2)). Only if the restoring force cannot compensate for an increasing stochasticity the probability for a jump over the potential barrier increases leading to an N-tipping event as sketched by the orange arrows. For completeness we also briefly discuss R-tipping in the supplementary material~\cite{a:supplNtip}, section \ref{sec:R tipping}. More detailed information can be found in the running text.
\end{tcolorbox}
state as shown in figure \ref{fig: control scheme} (b.2). In this case the fluctuations cannot overcome the potential barrier due to the restoring force in the potential landscape as schematically shown by the black and green arrows in Infobox 1, figure \ref{fig: control scheme} (b.3). 
However, the grid is also constructed out of components which live (partially) on a faster time scale assigned by orange (gray-orange), such as e.g. the unpredictable usage of end devices of private households and smaller machines of industries (apart from simple trends like the daily-periodic usage of electric light which are neglected in our scheme because of the much faster relevant time scales of power outages). 
\begin{wrapfigure}{RH!}{0.78\textwidth}
\begin{tcolorbox}[
width = \textwidth,
title=Infobox 2: N-tipping by a modified potential,
title filled=false,
colback=white!5!white,
colframe=black!75!black,
boxsep = 0.2mm,
fonttitle = \bfseries,
box align = base,
bottom = 2mm]
\floatbox[{\capbeside\thisfloatsetup{capbesideposition={center}, capbesidewidth = 0.63\textwidth}}]{figure}
{\hfill\includegraphics[width=0.33 \textwidth, left]{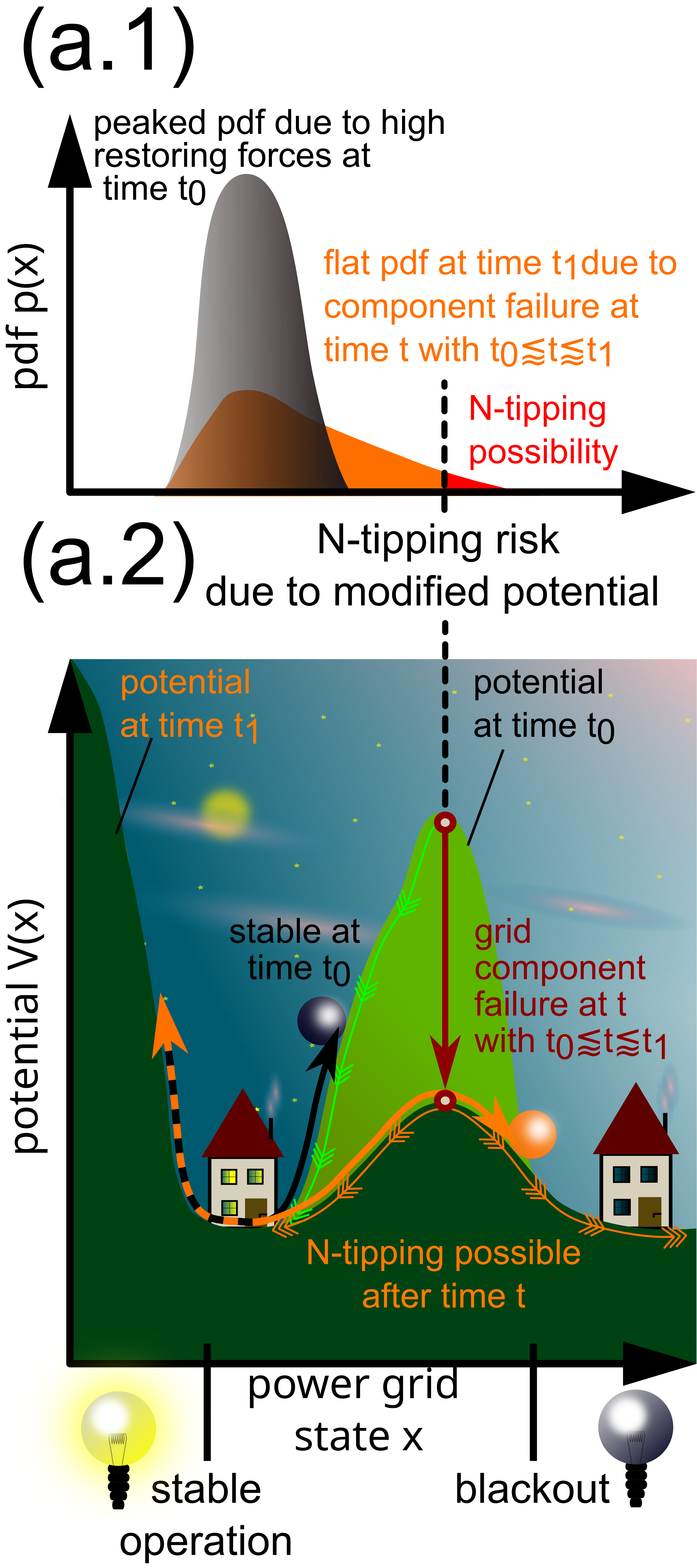}}
{\captionsetup{labelformat=empty}\caption[N-tipping due to a modified potential shape.]{\normalsize{\bfseries\textsf{Figure 2.}} N-tipping due to a modified potential shape. Additionally to the scenario sketched in Infobox1, figure \ref{fig: control scheme} (b.2, b.3), N-tipping can occur without strengthening of the noise, but under stationary noise level if the potential landscape is negatively affected, for example, due to grid component failures (cf. also Infobox 1, red arrows in figure \ref{fig: control scheme} (b.1)). In figure \ref{fig: mixed scenario scheme} it is sketched a possible scenario in which the high potential barrier of the light green potential at time $t_0$ shrinks down to the significantly lower barrier of the dark green potential with much flatter mountainsides at time $t_1$ by component failure at time $t$ with $t_0<t<t_1$. In consequence the restoring force is significantly decreased and crossing the weaker barrier after time $t$ is much more probable than before under the same noise level. This corresponds directly to a broadening of the pdf, similar to the one observed in Infobox 1, figure \ref{fig: control scheme} (b.2), but not because of changing contributions on a time scale of order~$\mathcal{O}(\tau_{\rm fast})$: Instead referring to figure \ref{fig: mixed scenario scheme} (a.1) the gray pdf corresponds to the localised stable operation state in the light green potential valley at time $t_0$ which becomes the much flatter orange pdf of the dark green potential at time $t_1$, although the noise level is stationary in this example. For completeness we also briefly discuss R-tipping in the supplementary material~\cite{a:supplNtip}, section \ref{sec:R tipping}.} \label{fig: mixed scenario scheme}}
\end{tcolorbox}
\end{wrapfigure}
\FloatBarrier
Since a sustainable adaptation of our power grids is indisputable, the influence of the faster time scales increases significantly by integrating modern technologies, e.g. wind and solar parks, from supergrids up to local ones and combined heat and power units (CHPs), solar cells, energy reservoir technology and e-mobility mostly on local grid scales. Basically, the increasing intrinsic stochasticity could --- in the hypothetical worst case --- flatten the grid state pdf $p(x)$ for a given grid infrastructure (a given potential landscape) as shown by the orange pdf in Infobox 1, figure \ref{fig: control scheme} (b.2). In such a scenario the intrinsic stochasticity can also lead to a power outage event by N-tipping, because the restoring force is not able to forbid the crossing of the barrier indicated by the orange arrow in Infobox 1, figure \ref{fig: control scheme} (b.3). In contrast to this scenario, Gloe et al.~\cite{a:Gloe2021} have discussed, how wind turbines' inertia and primary control capabilities could also help to stabilize the electrical infrastructure.
In order to accompany these theoretical considerations by practical ones, we briefly summarize the course of the power outage event of the North America Western Interconnection (NAWI) on 10th August 1996 data of which we will analyse in this article. The information sketched in this article is mainly taken from the approved \textit{Western Systems Coordinating Council Disturbance Report}~\cite{a:approvedReport} and the \textit{1996 - System Disturbances} report~\cite{o:NERC_report} from the North American Electric Reliability Council, but cf. also~\cite{o:NWPCC, o:Venkatasubramanian2003, ip:Venkatasubramanian2004, a:Kosterev1999, o:Hauer2010} and Appendices of the \textit{Preliminary Disturbance Report}~\cite{a:prelimReport}. At 14:01 PDT the \SI{500}{\kilo\hertz} Big Eddy-Ostrander line was opened by system protection because it flashed and grounded to a tree which led to several pre-outage line openings. These events led to an already significantly less resilient power grid state. Under these circumstances the tree-related high impedance fault~\cite{a:Bahador2018, a:Bahador2018b} (THIF) of the $\SI{500}{\kilo\volt}$ Keeler-Allston line had heavy impact on the grid stability: In the approved disturbance report~\cite{a:approvedReport} it is identified to be the key triggering event of the NAWI cascading failure on 10th August 1996. What follows was an interplay of control actions, component failures due to inadequate maintenance, human errors and finally a further cascade of various line openings which increased the reactive power in a way that the McNary units had to be removed from service between 15:47:40 and 15:49 PDT. This was the beginning of mild power oscillations in the transmission system which increased over time. As a direct consequence the entire grid was splitted into four isolated subgrid-islands: the Northern, the Northern California, the Southern and the Alberta island. This major power outage interrupted electric service to about \SI{7.5}{Mio.} customers the last of which got the electric service restored around 1:00 PDT at the 11th August 1996. 
Kosterev et al.~\cite{a:Kosterev1999} identified the power grid disturbance to be caused by a so-called \textit{small-signal instability}: 
The NAWI grid was known for damped interarea oscillations with a frequency around \SIrange{0.2}{0.3}{\hertz}. They were typical for the grid's topology which was vulnerable to some machine clusters oscillating against others through weak interconnections. If some control actions failed or the grid topology was unexpectedly modified, the damping could not stabilize the system anymore and the small signal disturbances were able to persist and grow leading to a power outage.\\
\begin{wrapfigure}{R}{0.75\textwidth}
\includegraphics[width = \textwidth]{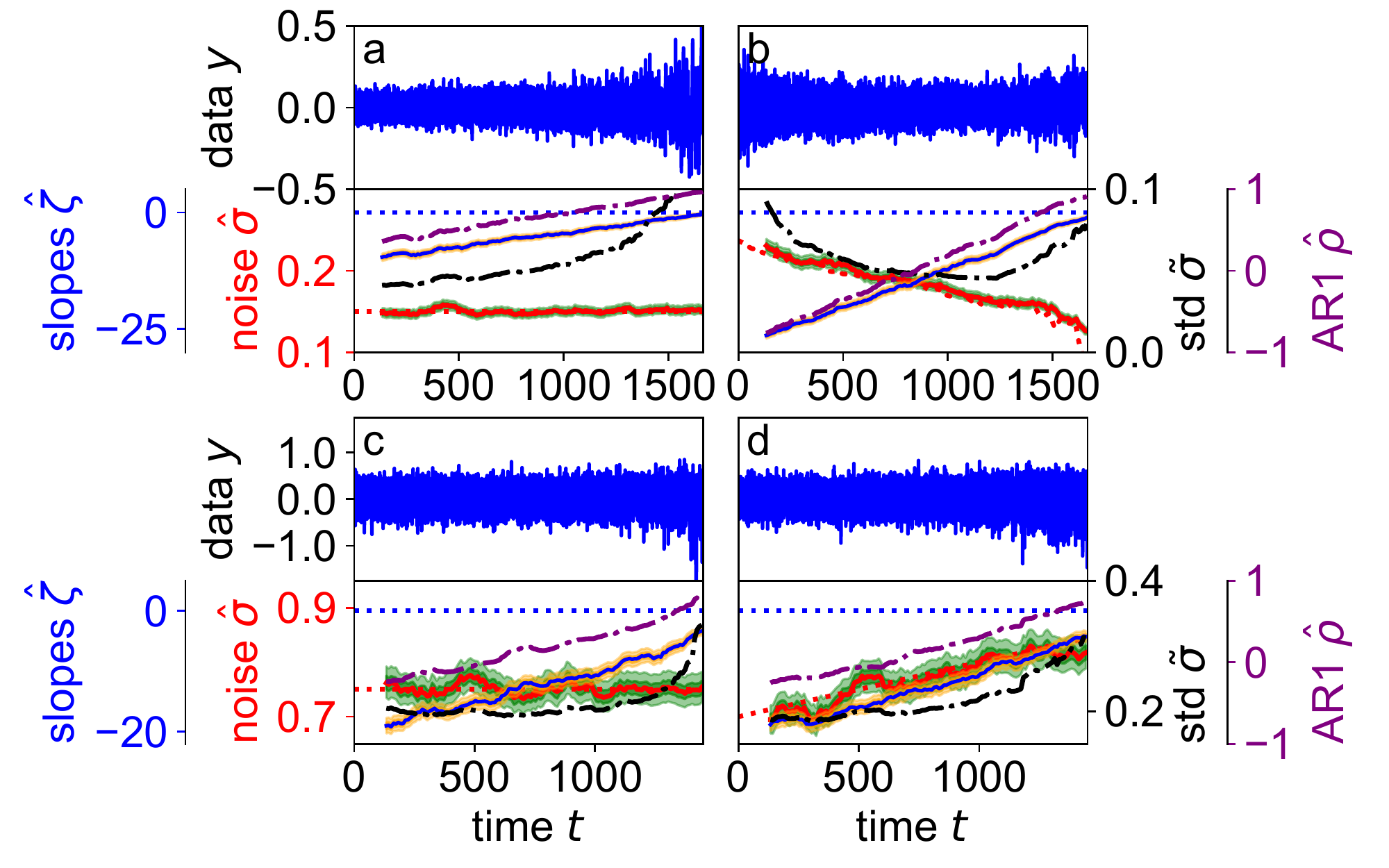}
\caption[Identify time series twins.]{
We give two examples of similar time series pairs (a,b) (similar in the second half) and (c,d) (similar over the whole range), but originate from pure B-tipping destabilization (a,c) or a combination of B-tipping and decreasing (b)/increasing (c) noise level. The analysis results, shown in the lower graphs, reveal the B-tipping candidates (a,c) by a drift slope estimate $\hat{\zeta}$ approaching zero (blue with orange credibility bands (CBs)) and the decreasing/increasing noise counterparts (b,d) by decreasing/increasing noise level estimates $\hat{\sigma}$ (red with green CBs) that match the true noise level (red dotted lines). In contrast to the BL-estimation results, the statistical leading indicators lag-one autocorrelation (AR1) $\hat{\rho}$ and standard deviation (std) $\tilde{\sigma}$ show almost the same fingerprints, namely positive trends, in three different destabilization scenarios (a,c,d) and thus do not provide information about the ongoing dynamical processes. Furthermore, they are not applicable in example (b) due to a convex std $\tilde{\sigma}$ curve that cannot be interpreted unambiguously in terms of a leading indicator. For more details we refer to the main text. 
}
\label{fig:JanusFace}
\end{wrapfigure}
\FloatBarrier 
The identified key factors that favored the outage are well-captured by our theoretical considerations. Even if the influence of renewable energy sources is rather negligible in the 1990s (cf. \textit{U.S. Energy System Factsheet} and \textit{Renewable Energy Factsheet} of the University of Michigan~\cite{o:USEnergySystemFactsheet, o:USRenewableFactsheet}) the report of the power outage event identifies two contributors on a fast uncontrollable time scale which in principle contribute to the diffusion $\underline{g}(\underline{x}(t) t)$ of the Langevin equation \ref{eq:langevin}. First, an unforeseen high energy demand in the private and public consumer section due to a heat wave and second, heavy energy exports due to favourable conditions of hydroelectric power generation (even if within established constraints). Both facts can contribute to a scenario as shown in Infobox 1, figure \ref{fig: control scheme} (b.2, b.3) or Infobox 2, figure \ref{fig: mixed scenario scheme}. Furthermore, the report points out several lapses in grid component maintenance, e.g. the omission of trimming trees near the lines or disregarding the significantly lower power supply from the Dalles hydro plant of which only five of 22 units were operating due to the well-known protection actions of salmon migration in this period. The latter key factors are directly contributing to the drift $\underline{h}(\underline{x}(t), t)$ of the Langevin equation \ref{eq:langevin} and could have modified the deterministic potential landscape negatively, basically following the red arrows in Infobox 1, figure \ref{fig: control scheme} (b.1).\\ 
Further reasons for preferring the BL-estimation rise from its capacity to distinguish more complex destabilisation scenarios as it would be possible with commonly used leading indicators for critical transitions. An illustration of why and how common statistical leading indicator candidates are limited compared to the alternative drift slope $\zeta$ is shown in figure \ref{fig:JanusFace} in which two example scenarios of critical transitions with and without changing noise level are analysed. 
The time series $y$ in the upper graphs of figure \ref{fig:JanusFace} (a,b) and (c,d) originate from models, exhibiting a pitchfork and a fold bifurcation, respectively.\\
The four time series have in common that they exhibit destabilizing control parameter shifts that finally lead to B-tipping. The two time series in figure \ref{fig:JanusFace} (a,c) undergo a pure B-tipping event by a rather simple linear control parameter shift, whereas the counterparts on the right in figure \ref{fig:JanusFace} (b,d) are additionally driven by decreasing (b) and increasing (d) noise levels. However, these distinctly more complex and more realistic scenarios generate rather similar time series Gemini in figure \ref{fig:JanusFace} (c,d) and a time series in \ref{fig:JanusFace} (b), whose second half looks similar to its time series twin in figure \ref{fig:JanusFace} (a). To accommodate the reading flow the model equations and a detailed description of the simulation procedure can be found in the section \ref{subsec:JanusExample}.
Only under rather restrictive conditions the common leading indicators are able to announce an uprising transition. For example, the parameter shift has to be slow and smooth to be followed by a statistical rolling window approach, the sampling rate needs to be high, the noise levels should be small, the choice of the observable is crucial and much more.~\cite{a:Perretti2012, a:gsell2016, a:Boerlijst2013, a:Hastings2010, a:Hessler2022} 
But more important is the fact that statistical measures as AR1 $\hat{\rho}$ or std $\tilde{\sigma}$ do not provide information about the origin of their raise, do not distinguish between pure B-tipping and more complex destabilization scenarios which for example, involve a variation of the internal noise and thus, can easily be ambiguous: The standard leading indicators AR1 $\hat{\rho}$ and std $\tilde{\sigma}$ have to increase \textit{both} at the same time if there is CSD prior to a B-tipping event, as e.g. visible in figure \ref{fig:JanusFace} (a). Unfortunately, the same system driven by decreasing noise level (approximately the red dotted line in the lower graph of figure \ref{fig:JanusFace} (b)) still destabilizes by B-tipping, but with an ambiguous convex shape of the std $\tilde{\sigma}$. The AR1 $\hat{\rho}$ and std $\tilde{\sigma}$ pair cannot be interpreted in terms of a B-tipping indicator and cannot even resolve the true decreasing noise level, i.e. a potentially lower N-tipping probability in case of a multi-stable system, in this example.\\
The AR1 $\hat{\rho}$ and std $\tilde{\sigma}$ computed on the time series Gemini of the model data in \ref{fig:JanusFace} (a,c,d) even result three times in rather the same signature, although the time series in $\tilde{\sigma}$ (d) is driven by increasing endogeneous disturbances in contrast to the time series in figure \ref{fig:JanusFace} (a,c) that are purely destabilizing by B-tipping. The AR1 $\hat{\rho}$ and std $\tilde{\sigma}$ do not provide information about the increasing noise level of the system in figure \ref{fig:JanusFace} (d). Note that in general it can have severe consequences if an increasing noise level during an approaching B-tipping event is not detected, because it causes a potentially higher N-tipping probability in multi-stable systems which can result into a transition significantly earlier than the time at which the actual B-tipping point would be reached. In this sense, the overall noise level has to be small enough to avoid N-tipping far before a B-tipping point if control and management decisions should be justified on AR1 $\hat{\rho}$ and std $\tilde{\sigma}$.\\
In contrast to these findings for the AR1 $\hat{\rho}$ and std $\hat{\sigma}$ the lower plots in our examples of figure \ref{fig:JanusFace} reveal the potential of the BL-estimation to detect the interplay of drift destabilization and noise level changes at the same time.
In the cases (a,c) the noise level (shown in red with green credibility bands (CBs)) stays constant, but the slope as leading indicator (in blue with orange CBs) approaches zero which indicates the ongoing B-tipping destabilization correctly. However, the examples of distinctly higher complexity in figure \ref{fig:JanusFace} (b,d) are more interesting. Whereas the AR1 $\hat{\rho}$ and the std $\tilde{\sigma}$ can only detect the B-tipping transition in figure \ref{fig:JanusFace} (d), the B-tipping risk in figure \ref{fig:JanusFace} (b) and (d) is correctly reflected by the increasing drift slope $\zeta$. Furthermore, in contrast to the AR1 $\hat{\rho}$ and the std $\tilde{\sigma}$ the decreasing/increasing noise level (marked by the red dotted lines), i.e. the changing N-tipping probability in multi-stable systems, is captured by the noise level estimates $\hat{\sigma}$ of the BL-estimation in the lower graphs of figure \ref{fig:JanusFace} (b,d), respectively. In conclusion, the introductory example confirms that the BL-estimation lends itself for a more robust and deeper analysis as it would be possible with the standard leading indicators and it could be applied both, for on-line controlling purposes as well as for an off-line scientific analysis of power outages with respect to the involved B-N-tipping risk factors evolving in time.\\
First, in section \ref{subsec: synthetic data} the method is tested against four synthetic time series of different systems to demonstrate the performance and limitations of the method. Second, two bus voltage frequency time series which cover periods before and after the power outage of the NAWI on 10th August 1996 are investigated with the BL-estimation to gain new insights into the complex interplay of the power grid's stability and varying noise influence in section \ref{subsec: power outage}. As suspected, apart from possible B-tipping the noise level seems to play an important role for the upcoming power outage. A previous study by Ehebrecht~\cite{mt:ehebrecht2017} supports our findings on the pre-outage time series by a related approach~\cite{a:Friedrich1997, a:Friedrich2000}. But in contrast to that study, we additionally contextualize our findings regarding the pre-outage and post-outage time series in close comparison to the real timeline of events and quantify uncertainties of the estimates. This leads to substantially deeper insights into the impact of certain events during the outage cascade and their modelling fingerprints. Furthermore, it widely confirms our hypothetical considerations in Infobox 1 and 2, figures \ref{fig: control scheme} and \ref{fig: mixed scenario scheme}. Moreover, we provide evidence for several mistakes in the derived time scale of the second post-outage time series used in previous studies~\cite{a:Hines2011, a:cotilla-sanchez12, mt:ehebrecht2017, a:grziwotz} which is most likely explainable due to sparse literature sources of low print quality (cf. supplementary material~\cite{a:supplNtip}, sections \ref{sec: metadata} and \ref{sec: time scale reconstruction}, for more details). The subsequent misinterpretation due to a wrong time scaling and time direction implies erroneously that cascading failures might be widely preceded by rather smoothly changing warning metrics built on the theory of CSD and B-tipping which might be surprising, considering the fact that grid components can turn out of service rather abruptly. Although our analysis supports basic theoretical considerations of the studies~\cite{a:Hines2011, a:cotilla-sanchez12, mt:ehebrecht2017, a:grziwotz}, it reveals a much more complex picture: The dynamics change rather abruptly related to real world grid modifications which can contribute in a stochastic and/or deterministic manner. Our results capture these state changes and account for the different contributing scales from stochastic to deterministic influence. In the end, a brief discussion of the results is given and complemented by suggestions of future research in section \ref{sec:discussion}.
\section{Results}
In section \ref{subsec: synthetic data} the Langevin approach is applied to four synthetic test cases, before the method is used to analyse the real world power outage data in section \ref{subsec: power outage}.
\subsection{Studies on Synthetic Data}\label{subsec: synthetic data}  
\begin{figure*}[!htp]
\subfigure{
\label{subfig: pitchfork data}
\includegraphics[width=0.24 \textwidth]{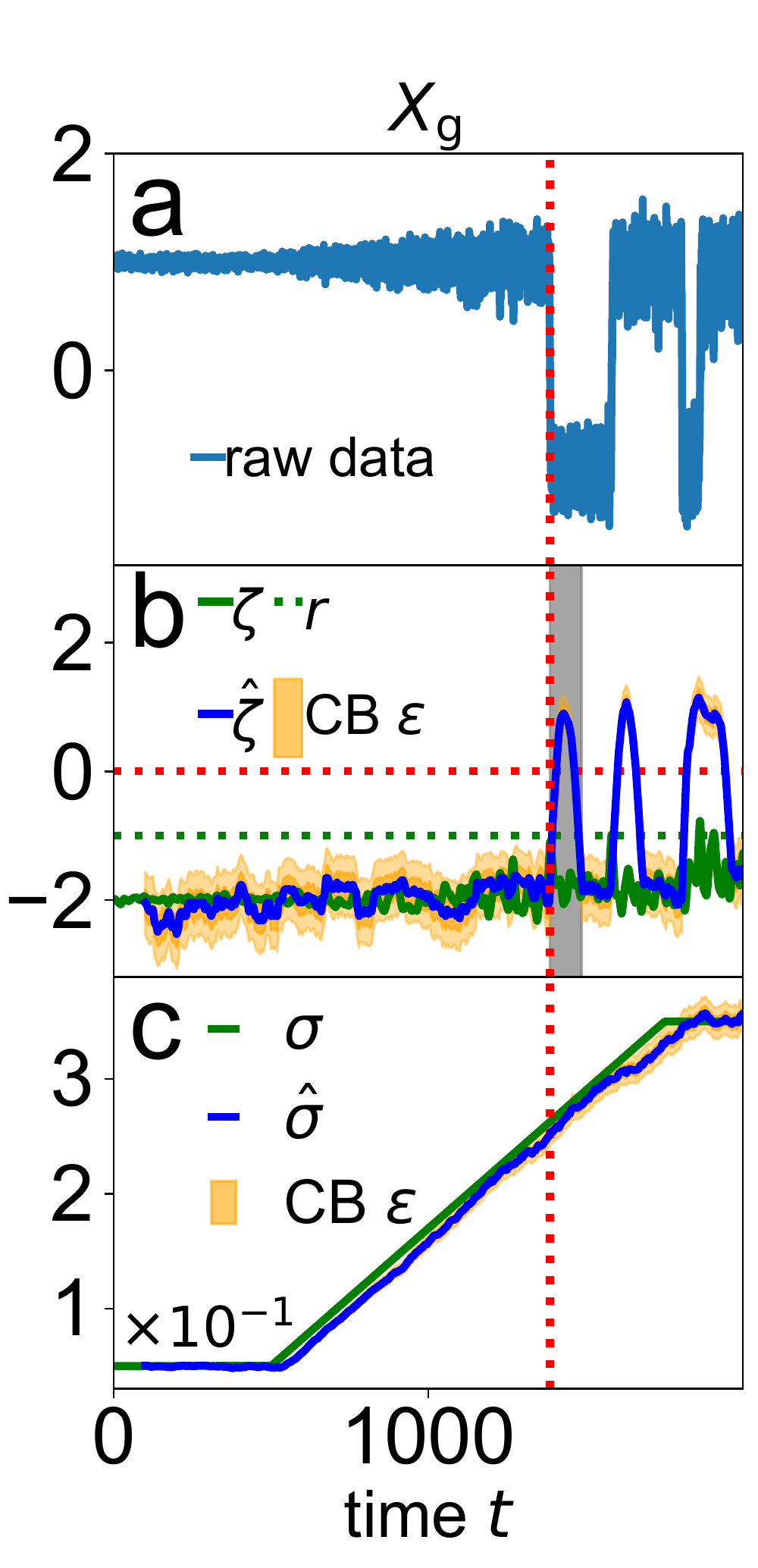}}
\subfigure{
\label{subfig: fold data}
\includegraphics[width=0.24 \textwidth]{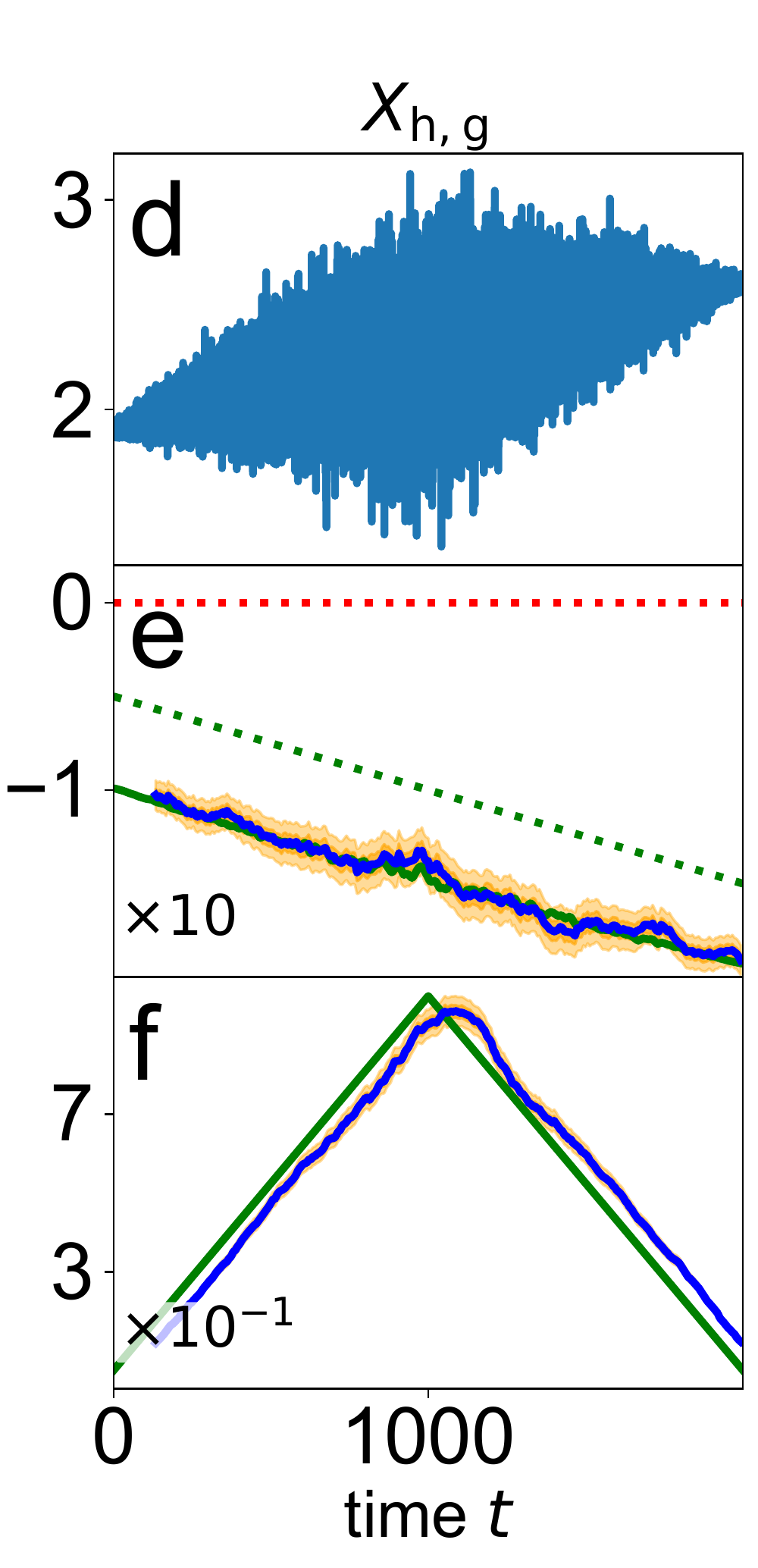}}
\subfigure{
\label{subfig: corr data}
\includegraphics[width=0.24 \textwidth]{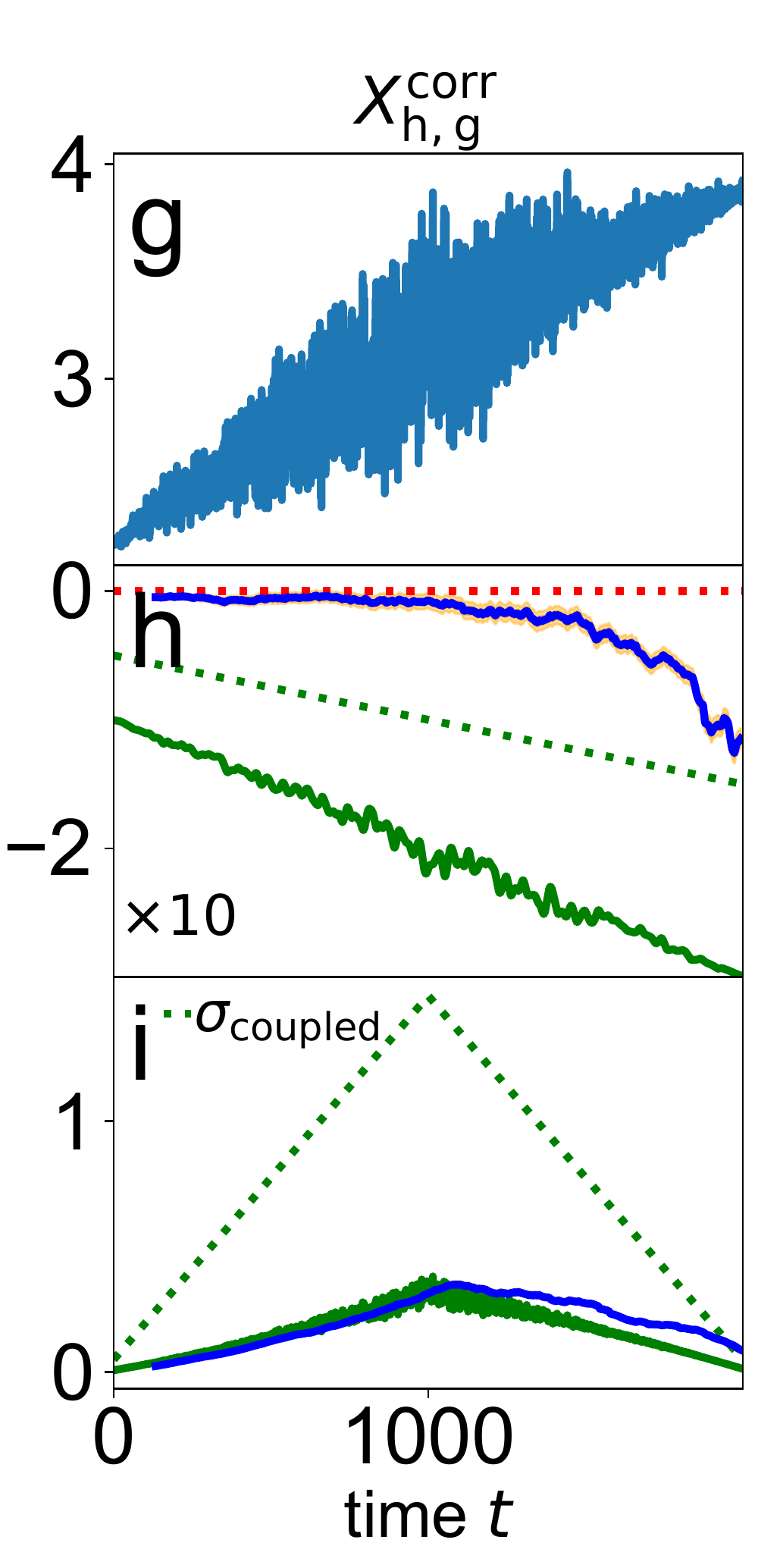}}
\subfigure{
\label{subfig: corr noise artefact}
\includegraphics[width=0.24 \textwidth]{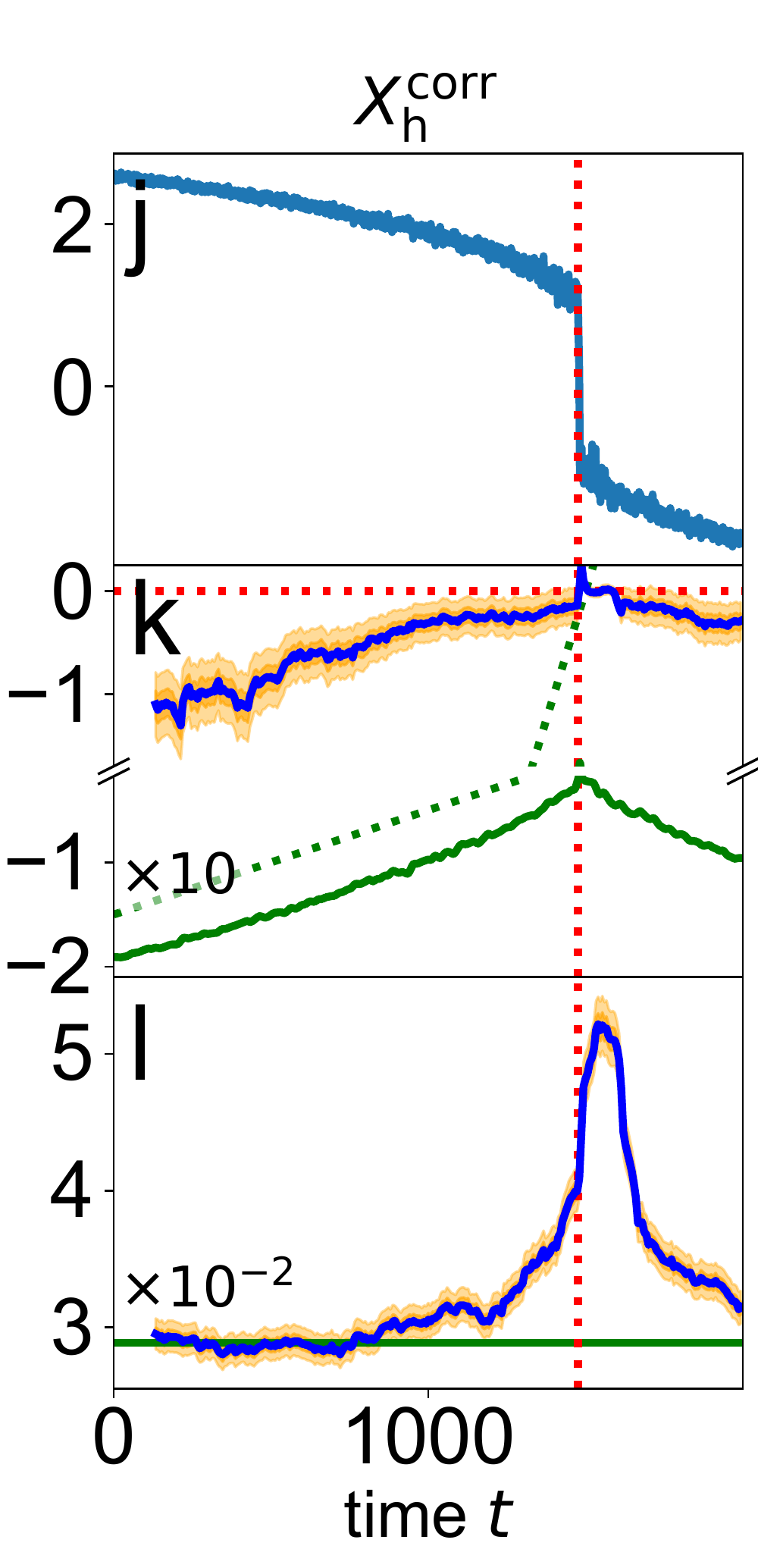}}
\caption[Tests on synthetic data.]{The results of the BL-estimation applied to the four test sets in (a, d, g, j). The red, dotted vertical lines in (a-c, j-l) indicate the approximate times when N-tipping into a flickering regime and B-tipping take place, respectively. The example datasets are analysed with the BL-estimation in windows of $2000$ points with a shift of $100$ points per window. In the examples (d,g,j), the data of each window is linearly detrended to account for the non-stationary trends in the mean.
(b, e, h, k) The drift slope estimates $\hat{\zeta}$ with the $\SI{1}{\percent}$ to $\SI{99}{\percent}$ and the $\SI{16}{\percent}$ to $\SI{84}{\percent}$ CBs are compared to the ground truth indicated by the green solid lines. The shift in time is due to the rolling window approach and ascribing the estimates to the last point of each window. Ascribing them to the mid point of each window, makes them match the true values almost perfectly (cf. supplementary information~\cite{a:supplNtip}, section \ref{sec: mid point aligned}, figure \ref{fig: no time lag}). For completeness the evolution of the control parameters is marked by the green dotted lines. The leading indicators $\hat{\zeta}$ mirror correctly the unchanged dynamics in (b) as well as the stabilizing/destabilizing dynamics in (e, h) and (k) by constant, negative and positive trends, respectively. The BL-estimation results of the Markovian examples are qualitatively and quantitatively correct. As expected the drift slope estimates are biased in the case of correlated noise which is confirmed by comparison to the analytical values in (h, k). Note the broken axis in (k). In (b) at time $t=1386.8$ the N-tipping causes artificial drift slope peaks with the width of one rolling time window as indicated by the grey-shaded area. (c, f, i, l) The evolution of the noise level $\sigma$ is compared to the estimates $\hat{\sigma}$ with the corresponding CBs. Apart from an expected time lag of the rolling window approach the original noise level is reconstructed in (c, f) and the first half of (i). In (i), the Markovian Langevin model assumptions become progressively less valid as the multiplicative coupling increases over time through $X_{\rm h,g}^{corr}$, and consequently, a precise quantitative reconstruction of the noise evolution fails in the second half of (i). Nevertheless, the qualitative behaviour of the noise level is preserved over the whole time range. The noise level estimates in (l) tend to show an artificial increase in the vicinity of a bifurcation, because the Markov assumption does not hold.}
\label{fig: synthetic results}
\end{figure*}

The following four datasets and two corresponding models are introduced in order to show that the method can track the noise level that eventually leads to N-tipping, during providing a resilience measure for B-tipping at the same time: The first example dataset is simulated with $40000$ data samples of the pitchfork model equations
\begin{align}\label{eq: pitchfork}
h_{\rm pf}(x) &=  - \nu \cdot x - x^3 \\
g_{\rm pf} (t) &= \sigma_{\rm pf}(t) = 
\begin{cases}
0.05 & \text{ for $t = [0, 500]$}, \\
2.4\cdot 10^{-4}\cdot t + 0.05 & \text{ for $t = (500, 1750]$}, \\
0.35  & \text{ for $t = (1750, 2000]$}
\end{cases}
\end{align}
with fixed control parameter $\nu = -1$. It has two stable states at $\pm 1$ and undergoes an N-tipping transition into a flickering state due to the increasing noise level. The example is called the \textit{N-tipping set} $X_{g}$. The index notation here and in the following indicates the terms that are varied in each time series.\\
The second system, called the simultaneous \textit{drift-diffusion-varying set} $X_{h,g}$ is governed by a fold model
\begin{align}\label{eq: fold model}
h_{\rm fold}(x) &= - r + x - x^3\\
g_{\rm fold}(t) &= \sigma_{\rm fold} (t) = 
\begin{cases} 9.5\cdot 10^{-4} \cdot t + 0.05 & \text{ for $t = [0, 1000]$} \\
-9.5\cdot 10^{-4} \cdot t + 1 & \text{ for $t = (1000, 2000]$}
\end{cases}
\end{align}
with linearly increasing and decreasing noise from $0.05$ to $1$ and vice versa in the first and second half of the simulated time interval $[0,2000]$. At the same time the control parameter $r$ decreases linearly from $-5$ to $-15$ over $30000$ samples over the whole time interval. The system would potentially undergo a fold bifurcation at $r_{\rm crit} = 0$ and thus, tends to stabilize with decreasing $r$ in the test case. Thus, the system could be imagined as a power grid in which diffusion is driven by variable noise as e.g. renewable energy sources, whilst the drift is continuously increasingly stabilized by controlling action. Naturally, a more sophisticated data-generating model, such as e.g. coupled Kuramoto oscillators with damping, results in some bias in a way proportional to the discreprancy between the model equations and the drift parameterization of the BL-estimation. For example, including damping of any form in the drift term undermines the direct correspondence of the BL-estimated noise level $\hat{\sigma}$ with the formally defined model noise level. Following the same basic mathematical reasoning, changes of a drift model damping term correspond to both changing drift slope and noise level estimates. However, the possible bias can in principle be avoided by introducing more suitable parameterizations of the BL-estimation in such cases. Note also that for monitoring uprising tipping risk factors the \textit{exact} estimates are much less important than their \textit{relative} time evolution which encodes the desired information.\\
Nevertheless, to balance the two Markovian study examples we introduce two systems with correlated noise to test the BL-estimation under non-Markovian conditions. The third example, called the \textit{correlated drift-diffusion varying set} $X_{h,g}^{\rm corr}$ is chosen to investigate the results of the method under the incorrect model assumptions of $\delta$-correlated noise: The model equations \ref{eq: pitchfork} are coupled to an Ornstein-Uhlenbeck-process (O-U-process) via
\begin{align}
\dot{x} &=  - \alpha \cdot x + x\cdot y - x^3 \\
\dot{y} &= - c \cdot y + \sigma_{\rm coupled}(t) \cdot {\rm d}W \label{eq:OU}\\
\sigma_{\rm coupled}(t) &=
\begin{cases}
1.45 \cdot 10^{-3} \cdot t + 0.05 & \text{ for $t = [0, 1000]$} \\
-1.45\cdot 10^{-3} \cdot t + 1.5 & \text{ for $t = (1000, 2000]$}
\end{cases}
\end{align}
with the control parameter $\alpha$ decreasing linearly from $-5$ to $-15$. The additive noise coefficient $\sigma_{\rm coupled}(t)$ of the O-U-process increases linearly in the range $[0.05, 1.5]$ in the first half of the simulation, before it decreases linearly to its start value $0.05$.\\
The last example, called the \textit{correlated B-tipping model} $X_{h}^{\rm corr}$ illustrates the fact that the noise estimates $\hat{\sigma}$ can tend to increase artificially for densely sampled non-Markovian datasets if the system approaches a bifurcation. Preliminary results suggest that this signature depends on the time scale separation of the processes $x$ and $y$. We observed this behaviour especially for a situation in which the slow observable was not measured, but only the fast one. The correlated B-tipping model $X_{h}^{\rm corr}$ is simulated via the model equation \ref{eq: fold model} coupled via the constant $q=0.5$ with an O-U-process (cf. equation \ref{eq:OU}) under constant noise influence $\sigma_{\rm coupled}(t) \equiv \sigma_{\rm coupled} = \sqrt{c}$ with $c=0.75$. A B-tipping destabilization is reached by a linear shift of the control parameter $r$ over the range $[-15,5]$ in the total simulated time interval. The corresponding simulations and the results of the analysis are presented in figure \ref{fig: synthetic results}, wherein control parameter trends of the models are shown as green dotted lines in (b, e, h, k). The example datasets are analysed with the BL-estimation in windows of $2000$ points with a shift of $100$ points per window. In the cases of the figures \ref{fig:
 synthetic results} (d,g,j), the data of each window is linearly detrended to account for the non-stationary trend in the mean. The $\SI{1}{\percent}$ to $\SI{99}{\percent}$ and $\SI{16}{\percent}$ to $\SI{84}{\percent}$ confidence bands (CBs) of the BL-estimates are represented by light and dark orange areas, respectively.\\
 
The N-tipping data $X_{g}$ is shown in figure \ref{fig: synthetic results} (a). The red dotted vertical line marks the time $t_{\text{N-tip}} \approx 1386.8$ at which the N-tipping transition into a flickering regime takes place. As a consequence of the stable dynamics of the system with $\nu \equiv -1$ the computed drift slope $\hat{\zeta}$ remains approximately constant up to time $t_{\text{N-tip}}$ after which it suddenly suggests a destabilisation by a sharp jump as visible in figure \ref{fig: synthetic results} (b). The beginning flickering causes artificial peaks of the leading indicator $\zeta$  that correspond to the sudden N-tipping transitions with the width of one rolling window as indicated by the gray-shaded area. The constant leading indicator level prior to time $t_{\text{N-tip}}$ matches the true drift slope values $\zeta$ marked by the green solid line and implies correctly no B-tipping transition.\\ 
But if it is not a B-tipping event, how can we identify the destabilizing mechanism? In this case estimating the slope $\zeta$ simultaneously with the noise level of the Langevin model yields valuable information about the ongoing phenomena: The deterministic dynamics do not change, but the uprising noise level is estimated well by the procedure as shown in figure \ref{fig: synthetic results} (c). The noise plateaus as well as the linear ramp of the noise level are followed precisely by the BL-estimation. The true noise level is marked by the green solid line, whereas its estimate is coloured in blue. Small deviations from the true values are expected because a rolling window approach always includes some time lag and the estimates are ascribed to the last point of each rolling window in the reasoning of an on-line resilience metric, i.e. the most recent BL-information are aligned with the present time stamp. Ascribing each estimate to the mid point of each window would make them match the true values almost perfectly (cf. supplementary information~\cite{a:supplNtip}, section \ref{sec: mid point aligned}, figure \ref{fig: no time lag}). Even though the noise level estimates cannot determine definitely whether an N-tipping event will occur or not, they provide valuable information about increasing noise and in direct consequence about a higher N-tipping probability in a system with multiple stable states.\\
In the previous N-tipping example $X_{g}$ the drift is unchanged. Is it possible to handle a more general situation where the noise level as well as the drift dynamics change simultaneously? To answer this question, the analysis is repeated for the drift-diffusion varying data $X_{h,g}$ as shown in figure \ref{fig: synthetic results} (d). The control parameter $r$ is modified linearly from $-5$ to $-15$ as visible in \ref{fig: synthetic results} (e). Note, that the drift slope $\zeta$ does not directly correspond to the control parameter $r$. If the control parameter $r$ drives the system towards destabilization, the drift slope $\zeta$ should approach zero and vice versa. More information can be found in section \ref{subsec:theory}. The slope estimates $\hat{\zeta}$ in the lower part of figure \ref{fig: synthetic results} (e) thus mirror correctly the stabilizing effects of the decrease of the control parameter $r$ by a negative trend and match the true drift slope values $\zeta$ marked by the green solid line. At the same time changes in the noise level are followed precisely apart from the above mentioned time lag because of a rolling window approach as illustrated by figure \ref{fig: synthetic results} (f). Briefly summarized, the two synthetic uncorrelated Markovian examples show that the BL-estimation can be a very advantageous tool to quantify tipping risk factors and to control systems with changing deterministic dynamics and noise level at the same time.\\
In principle the BL-estimation is limited to cases in which the general model ansatz of a Markovian Langevin equation holds (cf. section \ref{subsec:theory}). Nevertheless, it can yield useful results even in cases in which the model assumptions do not hold. This is an important point, because in a realistic case scenario a scientist cannot always determine for sure whether the assumption holds or not even if it is a reasonable model ansatz for many natural phenomena. A common violation of the parametric model is Non-Markovianity as present in the correlated drift-diffusion data $X_{ h,g}^{\rm corr}$ in figure \ref{fig: synthetic results} (g). The generating process of the $X_{ h,g}$-data is slightly modified in a way that violates the Langevin assumption by introducing sub-process noise over a multiplicatively coupled variable $y$. This leads to a non-Markovian process. 
The analysis results of this case are shown in \ref{fig: synthetic results} (h) and (i): The slope estimates $\hat{\zeta}$ still decrease and suggest correctly an increasing stability that is caused by the linearly decreasing control parameter $\alpha$, but as expected the estimates do not fit the ground truth (marked by the green solid line) anymore. In other words, the estimates are qualitatively --- but not quantitatively --- reasonable.
Interestingly, it turns out that the noise level estimates (blue line with orange CBs) in figure \ref{fig: synthetic results} (i) are still rather accurate in the example case especially in the first half of the simulated time range. The true noise level $\psi$ (green solid line) is determined following equation (14) in Willers and Kamps~\cite{a:Willers2021} divided by the time scale $\sqrt{\text{d}t}$ of the Wiener process. In our notation the general formula reads 
\begin{align}
    \psi = g_x(x) \cdot g_y(y) \cdot \text{d}t \label{eq:NonMarkovNoise}
\end{align}
with the coupling term $g_x(x)$ of the correlated noise process $y$ into $\dot{x}$ and the $y$-diffusion $g_y(y)$. This leads to the true noise level $\psi_{h,g}^{\rm corr} = x(t)\cdot \sigma_{\rm coupled}(t) \cdot {\rm d}t$ for the correlated drift-diffusion varying set $X_{h,g}^{\rm corr}$. The noise level $\sigma_{\rm coupled}(t)$ of the $y$-process is presented as green dotted line. With the general positive trend in the data $X_{h,g}^{\rm corr}$ the multiplicative coupling increases and the correlated process $y$ becomes more significant. That is probably the reason for the deviations of the noise level estimates $\hat{\sigma}$ in the second part of the time range.\\
At last, the analysis of the correlated B-tipping dataset $X_{h}^{\rm corr}$ is presented in the figures \ref{fig: synthetic results} (j-l). This example illustrates a more problematic estimation error that can occur if the Markov assumption is not sufficiently fulfilled in a densely sampled dataset. Even if the increasing drift slope estimates $\hat{\zeta}$ in figure \ref{fig: synthetic results} (k) suggest the approaching B-tipping event correctly, the noise level estimates $\hat{\sigma}$ exhibit an artificial positive trend in the vicinity of the bifurcation point. The constant true noise level $\psi_{h}^{\rm corr} = q\cdot \sqrt{c}\cdot \text{d}t$ derived from equation \ref{eq:NonMarkovNoise} is shown by the green solid line. It is worth to keep in mind that we have to carefully interpret a noise increase prior to a B-tipping event if the data is strongly non-Markovian. However, the general risk of a B-tipping event is also correctly detected in this non-Markovian example set.\\
In summary, the study of the BL-estimation on four synthetic datasets suggests that the method can yield more comprehensive insights into the dynamics combined with variable noise impact and is sufficiently robust against small violations of the model assumptions. The method provides an estimation of both resilience and noise level at the same time. Therefore, the BL-estimation lends itself to distinguish simple B-tipping from B-tipping with increasing N-tipping probability scenarios in contrast to common leading indicator candidates.

\subsection{North America Western Interconnection Power Outage on 10th August 1996}\label{subsec: power outage}
Keeping in mind the performance and robustness of the BL-estimation, demonstrated in section \ref{subsec: synthetic data}, we apply the algorithm to two bus voltage frequency time series $\omega_{\rm P}(t)$ and $\omega_{\rm R}(t)$ covering a historic major cascading failure, namely the NAWI blackout on 10th August 1996. As already sketched in section \ref{sec: intro}, the cascading failure led initially to a splitting of the NAWI into three islands around 15:48 o'clock and finally four islands, six minute later at 15:54 o'clock. It was a direct consequence of the McNary units tripping due to a system protection protocol, roughly around 15:47 o'clock. The system protection was forced to remove the power units from service owing to a preceding line opening cascade which started with tripping of the $\SI{500}{kV}$ Keeler-Allston line, again roughly five minutes earlier around 15:42 o'clock. This line opening is described as the key triggering event of the outage cascade. Altogether, the warm weather, heavy loads and inadequate tree-trimming favored the outage. In addition, the \textit{Bonneville Power Administration} (BPA) performed most of the grid component maintenance during summer because Northwest loads are typically lower during that time. Moreover, some hydroelectric power units were shut down to protect salmon migration. These circumstances already caused a weaker resilience, i.e. less ability of the grid control to react to sudden system component failures. The subsequent major power outage interrupted electric service to about \SI{7.5}{Mio.} customers the last of which got the electric service restored around 1:00 PDT at the 11th August 1996.~\cite{a:approvedReport, o:NERC_report, a:prelimReport}\\
We start our discussion with bus voltage frequency data $\omega_{\rm P}(t)$ from the pre-outage time interval in section \ref{subsubsec: before outage} and will conclude our considerations with the analysis of post-outage frequency data $\omega_{\rm R}(t)$ in section \ref{subsubsec: after outage}. The indices P and R denote the pre-outage and restoration intervals, respectively.\\
\subsubsection{The Pre-Outage Interval}\label{subsubsec: before outage}
In figure \ref{fig: power outage} (a-c), we present the BL-estimation results for the time series segment $\omega_{\rm P}(t)$ covering the entire timeline of events prior to the power outage. The analysed data $\delta [\omega_{\rm P}](t)$, shown in figure \ref{fig: power outage} (a), is the original frequency time series $\omega_{\rm P}(t)$ detrended by a Gaussian kernel smoothed version with kernel bandwidth $\SI{5}{\second}$ to transform the original data into a stationary version suitable for a statistical analysis. The BL-estimation is applied in windows containing $1000$ data points (i.e. $\SI{50}{\second}$) which are shifted by $100$ points (i.e. $\SI{5}{\second}$) to model the time evolution of the drift slope estimates $\hat{\zeta}$ and the noise level $\hat{\sigma}$, shown in figure \ref{fig: power outage} (b) and (c), respectively. The estimates are given by the blue lines with orange CBs corresponding to $\SIrange{16}{84}{\percent}$ and $\SIrange{1}{99}{\percent}$ percentiles as before. We interpret our results in close comparison to the official disturbance report~\cite{a:approvedReport} approved by the \textit{Western Systems(/Electricity) Coordinating Council} (WSCC at that time; now WECC). In the following presentation of the results note the high accuracy of the report's time stamps, since most of the grid components were satellite-synchronized.\\ 
The time series $\omega_{\rm P}(t)$ was provided by BPA via a \textit{Freedom of Information Act} (FOIA) request. Thanks to very cooperative correspondence we could clarify the time series' metadata (cf. the supplementary material~\cite{a:supplNtip}, section \ref{sec: metadata}, for more details on the data source): The data record is sampled with equidistant time steps $\Delta t=\SI{0.05}{\second}$ and was measured in Tacoma. The record starts at 15:29:40 o'clock. It ends at 15:48:54.95 o'clock which coincides approximately with the emergence of the first three islands, i.e.
\begin{itemize}
    \item the Northern Island which separated from California at 15:48:52.782 o'clock,
    \item the Californian part which separated into the Northern and Southern California Island at 15:48:54.765 o'clock.
\end{itemize}
 The historically significant cascading failure was triggered by the opening of the $\SI{500}{\kilo\volt}$ Keeler-Allston line due to a THIF~\cite{a:Bahador2018, a:Bahador2018b}. The line got into contact with a tree which led to flash over and subsequent tripping of the $\SI{500}{\kilo\volt}$ Keeler-Allston line at 15:42:03.139 o'clock which is marked by the end of the first red time interval in figure \ref{fig: power outage} (a-c) and led to almost simultaneous tripping of the Keeler-Pearl line. The sharply pronounced positive frequency deviation peak at that time mirrors the line tripping event. Based on our theoretical considerations about power grids, frequency dynamics and the Langevin model (cf. Infobox 1 and 2, figures \ref{fig: control scheme} and \ref{fig: mixed scenario scheme}) we expect to find variations in the drift slopes $\hat{\zeta}$ and the noise level $\hat{\sigma}$ as a fingerprint of the Keeler-Allston line tripping. And indeed we observe an anti-correlated trend of the drift slope and noise level estimates towards a new stable plateau starting even roughly two minutes prior to the actual recorded line tripping event of the approved disturbance report~\cite{a:approvedReport}. We highlight the approximate starting time of the change by the beginning of the first red time interval at 15:40:09.00 o'clock. At the same time there is a notable frequency dip in the original time series $\omega_{\rm P}(t)$ (cf. supplementary material~\cite{a:supplNtip}, section \ref{sec: frequency dip}) which can be an indicator of sharply increasing load in the grid. This dynamics' change preceding the actual line tripping is in good agreement to technical aspects of THIFs and inspires to two plausible explanations:
 \begin{enumerate}
     \item There is non-vanishing probability, that the contact between the Keeler-Allston line and the tree was established around 15:40:09.00 o'clock. The start time of the first red interval would therefore correspond to the beginning of the THIF. Since a THIF leads typically to a load increase on the directly affected line and the surrounding grid components, this idea is supported by the frequency dip at that time. The load increase might lead to higher impact of the fast time scale phenomena, i.e. increasing noise level $\hat{\sigma}$, stabilised by deterministic control mechanisms, i.e. a decreasing drift slope estimate $\hat{\zeta}$. Of course, we cannot deduce for sure causal relations between noise level and drift slope, even if the explanation seems to fit well into the overall picture. Deterministic changes could have also led to heavier impacts of fast scale phenomena. However, following the THIF scenario the grounded Keeler-Allston line led to a modified drift-diffusion dynamics corresponding to the new plateaus reached after 15:40:09.00 o'clock. After being grounded via vegetation it takes seconds up to several minutes until a line trips in case of a THIF.~\cite{a:Hou2015, a:THIF_timescale} Thus, the actual tripping event at 15:42:03.139 o'clock indeed falls into a reasonable time range. These findings suggest that THIFs could create fingerprints in the BL-estimation metrics based on bus voltage frequency data in some cases. That would be a very interesting starting point for future research projects on THIF detection.
     \item Though, it seems to be rather plausible that the changing dynamics are directly related to the actual Keeler-Allston line tripping by a THIF, we cannot deduce with certainty a causal relation between the state change and an uprising THIF. However, even if we assume the frequency dip at 15:40:09.00 o'clock not to be caused by the THIF, it probably corresponds to a sudden load increase by another reason. From an engineering point of view, a power grid load increase, especially in times of hot weather conditions, such as on 10th August 1996, favors the occurrence of THIFs: A power line heats up and elongates if load increases which leads to significantly stronger sagging of the line. In consequence, the distance between the power line and underlying vegetation shrinks and the threat of a THIF grows. This threat was additionally increased due to inadequate tree-trimming of BPA at that time. 
 \end{enumerate}
 Even though the exact circumstances cannot be deduced from the BL estimation, the method clearly identifies a significant change in the dynamics approximately two minutes than the actual line tripping event, and it appears highly likely to be related to the triggering event of the NAWI blackout on 10th August 1996.\\
What followed after the $\SI{500}{\kilo\volt}$ Keeler-Allston line flashed over, was a cascade of subsequent line trips over roughly $\SI{5}{\minute}$ due to THIFs, equipment failures and wrong manual operator decisions~\cite{a:approvedReport, a:prelimReport, o:NERC_report}. This period is resembled by the almost constant noise level plateau $\hat{\sigma}$ and the drift slope estimates $\hat{\zeta}$ fluctuating roughly around minus four between the red marked time intervals in figure \ref{fig: power outage} (a-c). The subsequent line trippings forced the McNary units to boost the reactive output to $\SI{494}{MVAR}$, before they started tripping at 15:47:36.00 o'clock because of technical issues with the excitation equipment which enforced actions of the system protection protocol. All four McNary units were lost at 15:47:52.00 o'clock. The McNary tripping corresponds to the second red interval in figure \ref{fig: power outage} (a-c). The BL-estimation identifies the lost reactive power of the McNary units by strongly increasing drift slopes $\hat{\zeta}$. This supports the drawn picture of the Langevin model for power grid frequency dynamics, since the McNary power units would approximately be determinstic grid components.\\
The loss of the McNary units is the crucial point in the cascading failure. The official report states that afterwards small frequency oscillations emerged which could not be damped anymore. Their increase led finally to the first island separation process at 15:48:52.782 o'clock. The BL-estimation mirrors this barely stable state by the new drift slope plateau $-1<\hat{\zeta}<0$ which is less stable than the state prior to the $\SI{500}{\kilo\volt}$ Keeler-Allston line flash over. Additionally, power grids are commonly modeled by coupled Kuramoto oscillators which contain a linear damping term.~\cite{ip:kuramoto1975, a:kuehn2019, a:rodrigues2016, a:tyloo2021} This damping can be interpreted in reality as control actions in order to stabilize the deterministic dynamics of the system. Actually, the approved disturbance report~\cite{a:approvedReport} states that several tools to damp frequency oscillations were not working at The Dalles and John Day due to plant control problems. This further underlines the barely stable plateau found in the drift slope estimates $\hat{\zeta}$. In the end, the drift slope estimates $\hat{\zeta}$ tend to increase more continuously which resembles the gradual destabilization due to increasing frequency oscillations. The unstable threshold zero is not reached, since the last data points are not evaluated due to the window shift of $100$ data points. However, the BL-estimation metrics yield much more detailed insights into the actual frequency dynamics and even indicate the uprising cascading failure roughly two minutes earlier than defined by the $\SI{500}{\kilo\volt}$ Keeler-Allston line's THIF.\\

\subsubsection{The Grid Restoration Interval}\label{subsubsec: after outage}
The second bus voltage frequency time series $\omega_{\rm R}(t)$, which we consider in this section, was originally extracted from four DIN A4 pages of an analog, printed frequency graph of the approved disturbance report~\cite{a:approvedReport} via an image processing software comparable to DigitSeis~\cite{a:Ishii2022}. The printed graph, originally provided by BPA for the disturbance report's task force, can be found in \textit{Exhibit 10} of the disturbance document~\cite{a:approvedReport}. In the articles~\cite{a:Hines2011, a:cotilla-sanchez12}, in which the digitized version of the frequency scan appeared for the first time, no absolute time stamps were reported. The studies used only relative times. After roughly $\SI{1.5}{a}$ of research and correspondence (cf. the supplementary material~\cite{a:supplNtip}, section \ref{sec: metadata}) to check the analog source of the digital time series $\omega_{\rm R}(t)$, we have strong reasons to build our analysis on a time axis that differs significantly from the originally reported one.
Based on the reasons stated in the supplementary material~\cite{a:supplNtip}, section \ref{sec: metadata}, we conclude that the investigated bus voltage frequency time series $\omega_{\rm R}(t)$ covers mostly the time \textit{after} the grid separation in four islands, incorporates roughly $\SI{20}{h}$ with an approximate time step resolution of $\Delta t=\SI{6.41}{s}$ and starts at 15:10:45 o'clock with an uncertainty of $\sigma_{t_{\rm s}}=\pm\SI{3}{\minute}$. The uncertainty in the range of a few minutes does not affect our conclusions, which operate on a scale of hours. Additional details regarding the approximate reconstruction of the correct absolute time stamps, time interval and the time step $\Delta t$ can be found in the supplementary material~\cite{a:supplNtip}, section \ref{sec: time scale reconstruction}.\\

\begin{figure*}
\subfigure{
\label{subfig: BV data}
\includegraphics[width=0.49\textwidth]{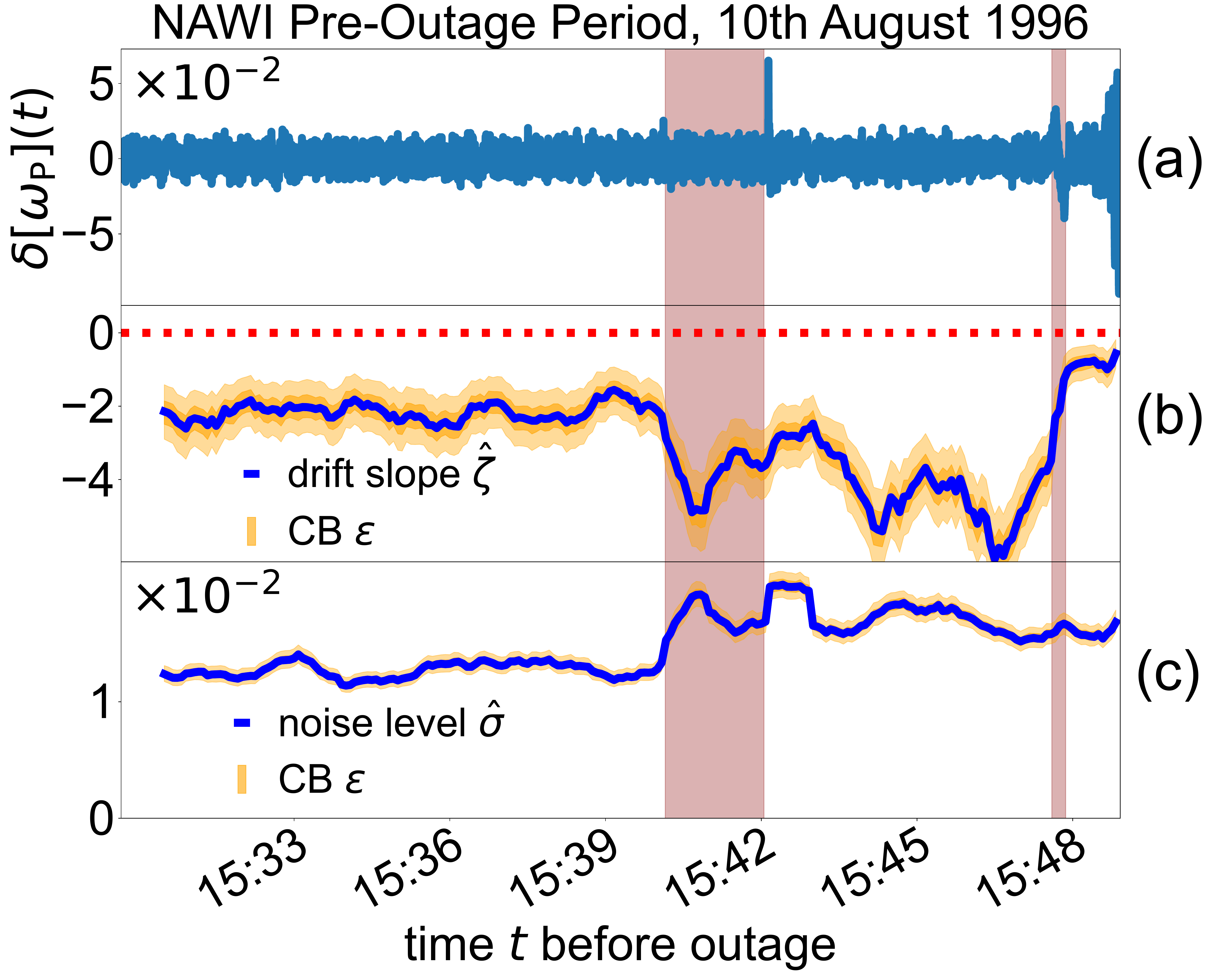}}
\subfigure{
\label{subfig: CD data}
\includegraphics[width=0.49\textwidth]{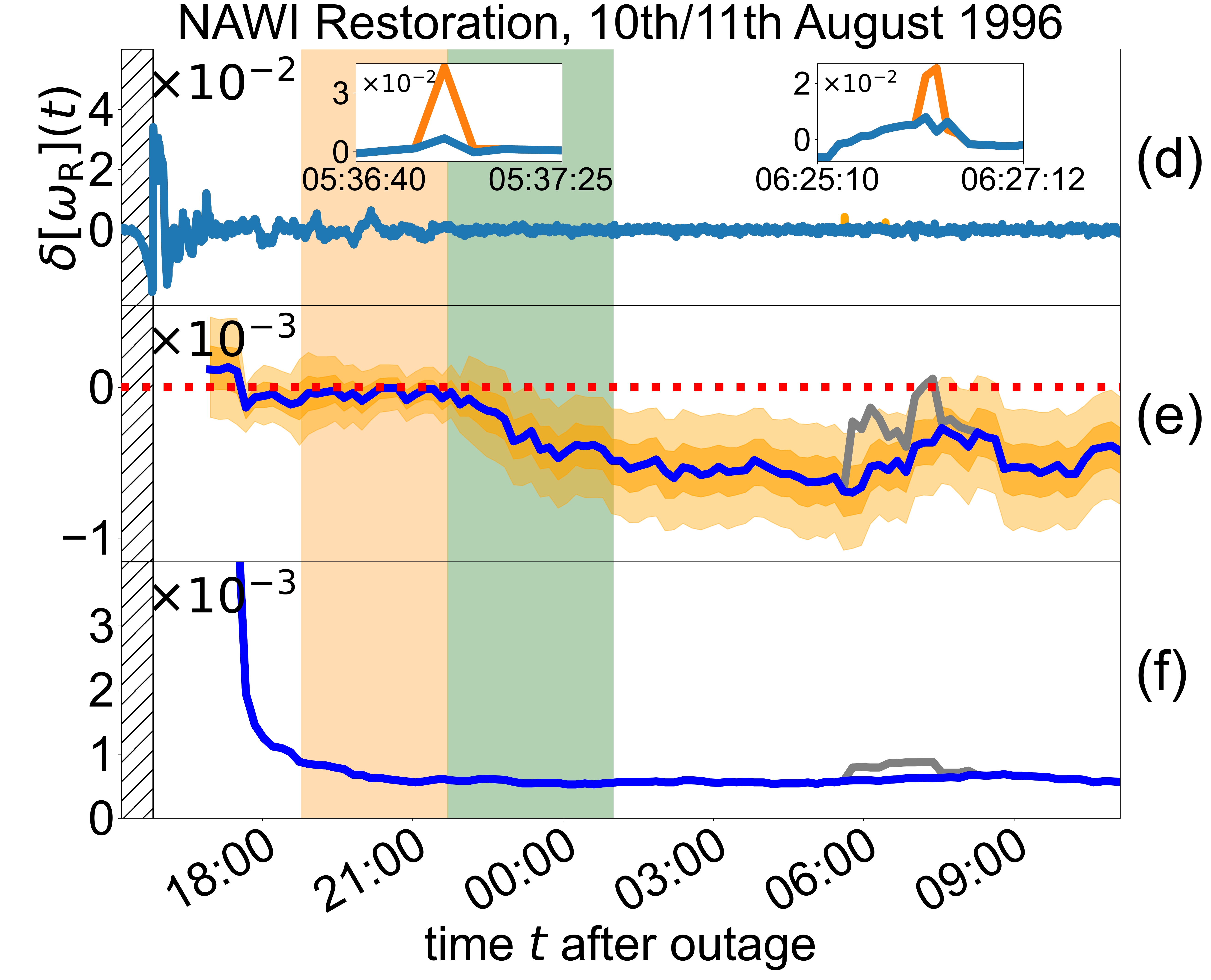}}
\caption[Power outage data analysis]{BL-estimation results for the pre-outage and restoration bus voltage frequency time series $\omega_{\rm P}(t)$ (a-c) and $\omega_{\rm R}(t)$ (d-f). The time axes are aligned with the time stamps of the approved disturbance report~\cite{a:approvedReport}. The end of the black-hatched interval falls together with the end of time series $\omega_{\rm P}(t)$. The shaded intervals follow a traffic light coloring from destabilizing to stabilizing. (a,d) The Gaussian kernel-detrended versions $\delta[\omega_{\rm P}](t)$ (kernel bandwidth $\sigma = \SI{5}{\second}$) and $\delta[\omega_{\rm R}](t)$ (bandwidth $\sigma = \SI{10.68}{\minute}$) are shown. They were used for the analysis. The insets in (d) show corrections (blue) of two artificial outliers (orange) (cf. section \ref{subsec: outlier correction} and the supplementary material~\cite{a:supplNtip}, section \ref{sec: outlier}) which cause as expected a discontinuous bias of the BL-estimates over one window length (cf. grey lines in (e,f)). (b) The drift slope estimates $\hat{\zeta}$ reveal mostly three different states over the pre-outage time interval. At the beginning of the red shaded interval at 15:40:09.00 o'clock the drift slope estimates $\hat{\zeta}$ converge to a new slightly more stable state roughly around $\hat{\zeta}\approx -4$, but with increased inter-window fluctuations. The state change could be triggered by a tree-to-line contact of the $\SI{500}{kV}$ Keeler-Allston line that tripped in consequence at 15:42:03.139 o'clock, marked by the end of the first red interval and easily observable due to the strongly pronounced peak in the time series $\delta[\omega_{\rm P}](t)$. If it is not directly caused by the tree-to-line contact, it still falls together with a dip in the original frequency time series (cf. supplementary material~\cite{a:supplNtip}, section \ref{sec: frequency dip}) at 15:40:09.00 o'clock which might indicate a sudden load increase in the system which favors tree-to-line faults. That suggests that the changing drift slope estimates $\hat{\zeta}$ decode an important permanent state change of the pre-outage region roughly two minutes earlier than the officially defined triggering key event of the historic cascading failure. The second red area covers the interval in which the four McNary stations were lost. Thereafter, increasing frequency oscillations led to the NAWI segregation into four islands with heavy and wide-spread blackouts in the end of the time series. The loss of the McNary units is resembled by rapidly increasing drift slopes $\hat{\zeta}$ reaching a new barely stable plateau. This observation is in good agreement with our theoretical considerations in Infobox 1, figure \ref{fig: control scheme}, in the sense that loss of a grid component can modify the potential landscape negatively. (c) The noise level estimates $\hat{\sigma}$ change anti-correlated to the drift slopes $\hat{\zeta}$ at the beginning of the first red interval which implies higher impact of fast scale phenomena in the grid, although the potential landscape was slightly more narrow than before. This hints to a strongly increasing impact of fast scale phenomena, since the more narrow potential cannot damp the noise level increase. The loss of the McNary units in the second red interval has no influence on the noise level which supports our model reasoning by similar arguments as before. (e) The first drift slope estimates $\hat{\zeta}>0$ are not trustworthy, since the windows incorporate pre-outage data. In particular, the estimates are sensitive to the dip and pronounced peak of the beginning outage at 15:48 o'clock. The drift slope suggests a barely stable grid over thet orange shaded key restoration interval. We define it from re-synchronization of the last island at 18:47 o'clock until 21:42 o'clock when the full load of the system was restored. The indicated weak stability makes sense, since most of the grid components are only step-wise recovered during the marked interval, some of them even much later, i.e. over the next days until the 16th August 1996. After the load restoration, the system approaches steadily the physical pre-outage configuration. At 01:00 o'clock in the morning of 11th August 1996 the last customers were supplied with electric energy again which corresponds to the end of the green shaded period. The drift slope indicates a new stabilization plateau. (f) The first noise level estimates $\hat{\sigma}$ are not trustworthy analogously to (e). Afterwards the noise level exhibits a negative transient with ongoing restoration of the grid which illustrates the steepening of the potential that traps the grid state in the potential valley's stable operation fixed point according to Infobox 1, figure \ref{fig: control scheme} (b.1). For more detailed information we refer to the main text.}
\label{fig: power outage}
\end{figure*}

Similar to the technical details in section \ref{subsubsec: before outage}, we detrend the original frequency time series $\omega_{\rm R}(t)$ by a slow trend version obtained by Gaussian kernel smoothing with the bandwidth $\sigma = \SI{10.68}{\minute}$ and obtain the detrended frequency time series $\delta [\omega_{\rm R}](t)$, shown in figure \ref{fig: power outage} (d). For the analysis of the detrended time series version $\delta [\omega_{\rm R}](t)$ we use time windows of $1000$ data points (i.e. $\SI{1.78}{h}$) with a shift of $100$ points (i.e. $\SI{10.67}{\minute}$).\\
Following the interpretation of the metadata from the supplementary material~\cite{a:supplNtip}, the black-hatched time interval in figure \ref{fig: power outage} (d-f) covers only pre-outage data. The end of the black-hatched interval matches with the end of the high resolution frequency data in figure \ref{fig: power outage} (a) at 15:48:54.95 o'clock. The first three islands which emerged from 15:48:52.782 to 15:48:54.765 o'clock correlate with the first saw tooth-shaped peak. Roughly six minutes later, the final system separation state was reached, when Alberta segregated from the Northern Island, creating the Alberta Island at 15:54:00.00 o'clock. The separation of the Alberta Island falls together with the second saw tooth-shaped peak, which is admittedly difficult to see due to the time scale resolution.\\
The Alberta Island was the last area that insulated from the NAWI, but also the first one that re-synchronized with the Northern Island, namely at 16:29 o'clock. The full load in the Alberta area was already restored at 17:39 o'clock which, at first glance, seems to coincide with the zero-crossing of the drift slope estimates $\hat{\zeta}$ and the rapid transient to stable noise level estimates $\hat{\sigma}$ in figure \ref{fig: power outage} (e) and (f), respectively. However, the estimates are not trustworthy, since the first windows incorporate pre-outage and post-outage data. The zero-crossing of the drift slope estimates coincides with the exclusion of the first dip and pronounced first saw tooth-shaped peak around 15:48 o'clock. In that sense the zero-crossing is sensitive to the rolling window size because of non-stationary data (i.e. mixed pre-outage and post-outage data) and does not reflect the real world event of load restoration in Alberta after its re-synchronization.\\
Only shortly after 21:42 o'clock when the full load was installed again by recovering the \textit{Metropolitan Water District/Southern}, the drift slope estimates $\hat{\zeta}$ imply further stabilization by a negative trend towards a stable plateau around $\hat{\zeta}\approx 5\cdot 10^{-4}$ in the green-shaded time interval. During this interval the last customers were connected to the grid again until 01:00:00 o'clock~\cite{o:NERC_report}. This is in good agreement to our theoretical model considerations, since after that time the grid resembles more and more the early pre-outage state. Analogously, the rapid noise level transient prior to the orange-shaded key restoration interval, slows down within the orange shaded interval in order to reach a stable plateau in the green shaded period. The complete restoration of the NAWI took several days as already mentioned in some examples above. \\
The scaling differences of the BL-estimates from the frequency time series $\delta [\omega_{\rm P}](t)$ and $\delta [\omega_{\rm R}](t)$ in figure (a-c) and (d-f), respectively, are caused by the significantly different time resolution and quality of the data. We want to conclude with a few remarks on that. In principle, generating the digitized version of the frequency time series $\omega_{\rm R}(t)$ of the restoration interval from a printed graph via an image processing tool, leads to reasonable approximated data for our analysis. However, we noticed some outliers in the extracted raw data. We show the raw data in orange in figure \ref{fig: power outage} (d) with the two most prominent outliers in the insets. The effect of these outliers is demonstrated by the grey drift slopes $\hat{\zeta}$ and noise level estimates $\hat{\sigma}$ in the figures \ref{fig: power outage} (e, d). Essentially, they were concomitant with discontinuous jumps in the estimates to plateaus of approximately one window length. With access to a scan of the original printed frequency time series we could identify these outliers to be artefacts of the data generation procedure. Basically, the image processing algorithm that extracted the digitized time series from the printed version misinterpreted grid lines of the millimeter paper to be actual data points (cf. the supplementary material~\cite{a:supplNtip}, section \ref{sec: outlier}, figure \ref{fig: outlier} for more details). In light of this, the discontinuous jumps in the BL-estimates make sense, since the outliers are completely independent from the assumed stationary data distributions in the individual windows and has to create a bias for that reason. The unbiased BL-estimates, presented in blue with orange CBs in the figures \ref{fig: power outage} (e, f), are accordingly computed on a corrected version of the frequency time series, shown in blue in figure \ref{fig: power outage} (d). To this end, we used a thinned version of the data and replaced the two most prominent outliers systematically by random values to diminish their influence without making assumptions about the underlying frequency dynamics. For more details concerning the outlier correction procedure see section \ref{subsec: outlier correction}. After that, the BL-estimates of the outlier-corrected dataset show only a weak increase over one window length in the corrected intervals. This is an expected result, since the correction values were drawn from a distribution estimated in a symmetric $\epsilon$-environment around the outlier sequence. Therefore, they correct for the strong outlier magnitude, but do not encode information about the frequency dynamics in the corrected intervals or in other words, the increments between the random values do not resemble the actual frequency dynamics. All in all, we assume that the power grid remained in a similarly resilient state from 01:00 o'clock until 11:00 o'clock. 

\section{Discussion and Outlook}\label{sec:discussion}
The BL-estimation procedure was shortly introduced with the drift slope $\hat{\zeta}$ as resilience measure and the noise level estimate $\hat{\sigma}$ that is related to the N-tipping probability in multi-stable systems. In an introductory example both measures were compared to the common leading indicator candidates AR1 $\hat{\rho}$ and std $\tilde{\sigma}$. The time series of the introductory example, that look similar, but pass different mechanisms to destabilization, can be explained as an interplay of drift and diffusion evolution by the BL-estimation procedure, whereas the AR1 $\hat{\rho}$ and the std $\tilde{\sigma}$ show ambiguous behaviour or are not able to resolve the different underlying mechanisms.\\
After this comparison, the BL-estimation is applied to four synthetic time series under variable conditions to investigate its performance. The method is able to follow precisely the changing drift dynamics as well as it can mirror the noise evolution while the assumption of a Langevin model holds. Nonetheless, the qualitative evolution of the stability and the noise level is also possible if the assumption is violated which is an important implication for many real world applications.\\
The method is used to analyse two bus voltage frequency time series from the power outage event in the NAWI on 10th August 1996 which represent basically a pre-outage and a post-outage frequency record. A detailed comparison of the BL-estimation results to the real timeline of events supports our theoretical ideas about power grid frequency dynamics and the Langevin model to large extent (cf. section \ref{sec: intro}). Our results disprove the simple idea of smooth changes of early warning metrics based on CSD and the theory of B-tipping, even if the basic assumptions are still supported. Instead of rather continuous changes in the frequency dynamics, i.e. the power grid state, we observe fast variations into states of modified deterministic resilience and stochastic influences which are closely interrelated to abrupt real events like line openings, loss of power units or failures of other grid components. These state changes are almost discontinuous in nature. However, the alternating states are mirrored by BL-estimates approaching alternating constant values. These findings fit to common intuition of cascading failures which are not thought to be continuous, but rather discrete series of destabilizing events. Nevertheless, the analysed example shows that they can lead to operating conditions which trigger indeed to more continuous destabilization, i.e. a lack of damping options that favored increasing frequency oscillations for approximately one minute prior to the first formation of grid islands. Even if a skilled power system controller might be able to extract already much of the important information about changes in the grid state from the raw frequency time series, the distinction of fast and slow dynamics and the persistence of changes in the BL-estimates add valuable information. In particular, the BL-estimation identifies a significant and persistent change in the frequency dynamics two minutes before the outage's key triggering event, i.e. the tripping of the $\SI{500}{kV}$ Keeler-Allston line, took place. Even if a trained system controller might have noticed a short dip of the frequency at the time suggested by the BL-estimation, the dip does not yield information about the persistence of the system's state. In the considered outage the newly reached stable constant values persist for roughly six minutes, until the loss of the McNary units changes the state again. In both cases the results give a more comprehensive picture of the interplay of changing deterministic dynamics and noise influence than it would be possible with previously applied leading indicator candidates. The BL-estimation is provided in the open-source Python package \textit{antiCPy}~\cite{url:GitHessler2021, url:DocsHessler2021} and is an alternative candidate for various stability analysis purposes. Application of the implemented algorithms to other systems could improve the understanding of its scope and limits. For example, it could turn out to be very interesting to adapt the drift and diffusion parameterization based on partial system knowledge and specific established models to investigate if and how quantitative estimation biases can be avoided. Furthermore, we think about including prior information about multi-stability, i.e. restrict the parameter space to those combinations which realise a double-well potential. In that way it might be possible to adopt the formalism of Kramers's rate estimation without having access to data of \textit{both} minima of the double-well potential. Given that we assume a priori a bi-stable system might enable us to extract the needed potential barrier height to compute the Kramers's rate if the noise is strong enough to resolve the inflexion point of only one of the potentiatl valleys.\\
Yet another interesting pathway of future research could be rate-dependent tipping, since in principle the slope of the drift slope estimates $\hat{\zeta}$ relates to the rate of change of a drift control parameter which in turn drives R-tipping if the rate of change is sufficiently high. However, it remains an open question whether the BL-estimation could be adapted to resolve the (by definition) fast R-tipping rates of change --- keeping in mind the rolling time window delay and the necessary minimum amount of data per window. And even if these issues can be solved, the relation between the slope of the drift slope estimates and the system-dependent critical rate of change for R-tipping are normally unknown. Nevertheless, it may be possible to manually determine the critical thresholds for specific systems at hand, but this is a complex research topic on its own.\\  
Even if the performed BL-estimation analysis of the power outage is able to distinguish between changes in the resilience of the power grid and the internal noise level that may push it across the separatrix between two alternative stable states, it leads to the novel question how these changes may or may not influence each other. Furthermore, up to now the procedure tends to give slightly increasing noise estimates prior to a bifurcation if the Markov condition is not sufficiently fulfilled. Preliminary results suggest that this behaviour is related to analyses of observables on the fast time scale in which the actual relevant slow observable is not measured directly. However, these are only preliminary hypotheses. Therefore, in future research we focus on improvements of the estimation procedure also for Non-Markovian time series. 

\section{Data and software availability and supplementary information}
The simulated and empirical data as well as supplementary information can be found in the github repository \url{https://github.com/MartinHessler/Disentangling_Tipping_Types}. The open-source package \textit{antiCPy} to perform the BL-estimation can be found at \url{https://github.com/MartinHessler/antiCPy} under a \textit{GNU General Public License v3.0} and is documented at \url{https://anticpy.readthedocs.io}.

\section*{Materials and Methods}\label{sec:method}
\section{Simulation details of the introductory examples}
\label{subsec:JanusExample}
\begin{wrapfigure}[21]{r}{0.33\textwidth}
    \includegraphics[width = \textwidth]{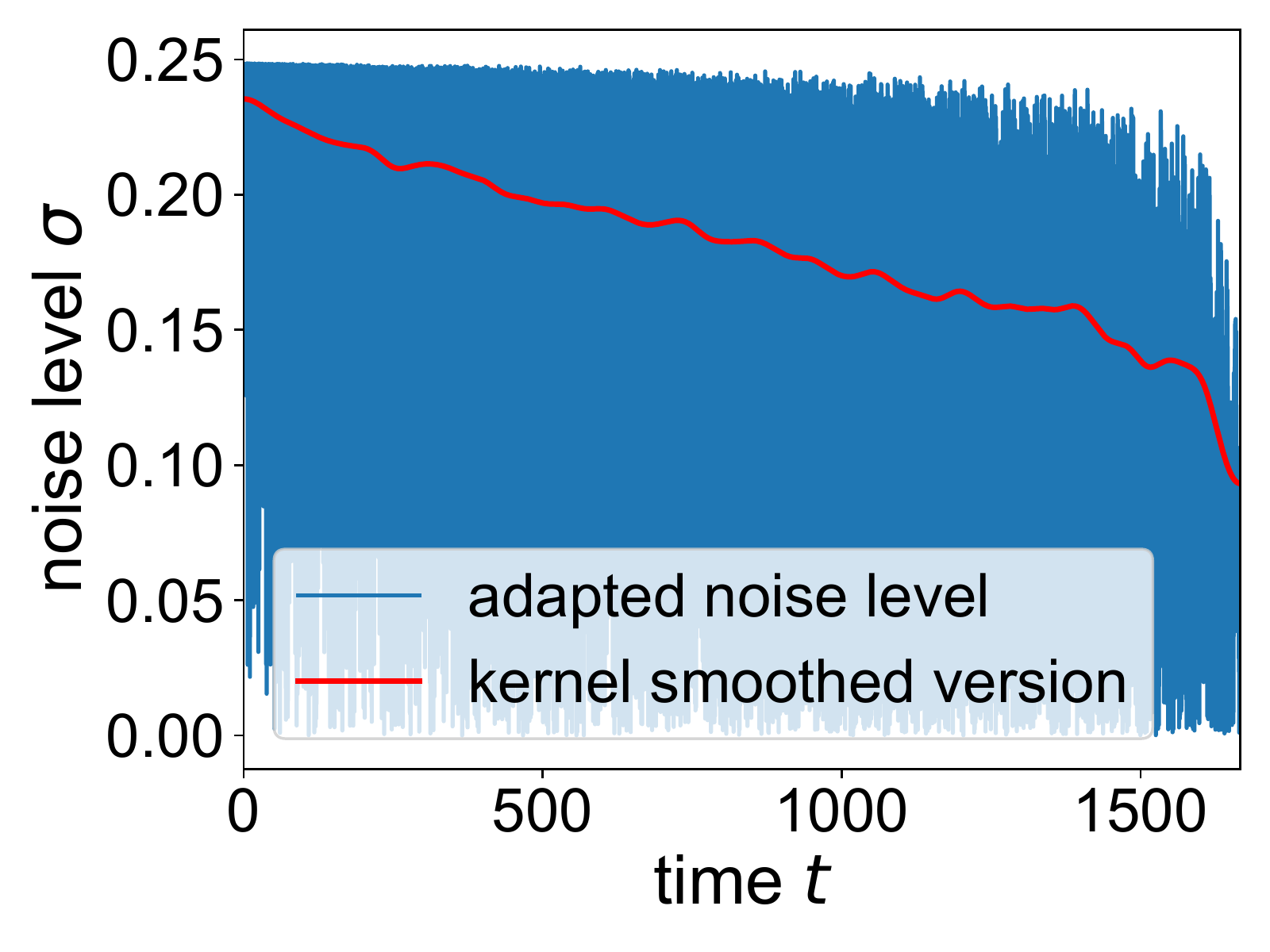}
    \caption[Adapted noise level of the pitchfork introductory example.]
    {The computed noise level results $g(x)$ are shown for the pitchfork model with linearly shifted control parameter from $[-28,5]$ if the noise $g(x)$ is matched to the data distribution in equation \ref{eq:distribution} of the time series in figure \ref{fig:JanusFace} (a). A smoothed version via a Gaussian kernel with bandwidth of $\sigma_{\rm kernel} = 325$ points (i.e. $\sigma_{\rm kernel} = \SI{21.6}{a\text{.}u\text{.}}$) is also shown and corresponds to the shaded red dotted line in the lower graph of figure \ref{fig:JanusFace} (b).}
    \label{fig:noiselevel}
\end{wrapfigure}

For the sake of readability the details of the simulations of the pitchfork and the fold model shown and analysed in figure \ref{fig:JanusFace} are summarized in the following. The pitchfork models of figure \ref{fig:JanusFace} (a,b) follow the model equations \ref{eq: pitchfork}. In the purely B-tipping case (a) the control parameter is linearly shifted in the range $[-10, 2]$ and $30000$ data samples are simulated over the time interval $[0,2000]$ after a burn-in period of the initial $500$ samples. The noise level is chosen constant as $\sigma = 0.15$. In order to create the similar dataset with decreasing noise level shown in figure \ref{fig:JanusFace} (b) the pitchfork model is simulated with a control parameter linearly shifted in the different range $[-28,5]$ and adaptation of the noise level corresponding to the distribution~\cite{b:Haken2004}
\begin{align}
    p(x) \propto \exp \left(\frac{-2V}{Q}\right)
\end{align}
of the corresponding Fokker-Planck equation of the time series (a), with the potential
\begin{align}\label{eq:distribution}
    V(x) = -\int_{x_0}^x h(x) {\rm d}x
\end{align}
 and the diffusion coefficient 
 \begin{align}
     Q = \frac{1}{2} g^2(x),
 \end{align}
 where $h(x)$ and $g(x)$ are the drift and diffusion, respectively, of the Langevin equation \ref{eq:langevin}. The very noisy results of this noise level estimates $g(x)$ and a version that is smoothed via a Gaussian kernel with a kernel width of $\sigma_{\rm kernel} = 325$ points (i.e. $\sigma_{\rm kernel} = \SI{21.6}{a\text{.}u\text{.}}$) are shown as blue-solid and red-dotted lines, respectively, in figure \ref{fig:noiselevel}. The smoothed version is also shown as a red dotted line in figure \ref{fig:JanusFace} (b) to guide the eye. The pitchfork models are cut for times in which the control parameter $\nu \leq 0$.\\
The fold model is computed via the model equations \ref{eq: fold model}. The purely B-tipping version of figure \ref{fig:JanusFace} (c) is calculated with a linearly shifted control parameter in the range $[-15,5]$ and a noise level of $\sigma = 0.75$ over the time range $[0,2000]$ with $30000$ data samples. The time series twin in figure \ref{fig:JanusFace} (d) is simulated similarly, but with a linearly shifting noise level shift in the range $[0.7,0.9]$. The fold models are cut for times in which the control parameter $r\geq -0.5$. The cut versions are detrended by subtracting a Gaussian kernel smoothing with bandwidth $\sigma_{\rm kernel} = 1500$ points (i.e. $\sigma_{\rm kernel} = \SI{100}{a\text{.}u\text{.}}$).\\
\section{Numerical method}\label{subsec:theory}
The Bayesian Langevin estimation procedure, explained in the following, is applied in rolling windows of the time series data. Similar to the idea in Carpenter and Brock~\cite{a:carpenter11} the observed signal in each window is modelled by the stochastic differential Langevin equation~\cite{b:Kloeden1992}
\begin{equation}
\dot{x}(x,t) = h(x(t),t) + g(x(t), t) \Gamma (t),
\end{equation}
where $h(x,t)$ is the nonlinear deterministic part, the so-called drift and $g(x(t), t)$ is the diffusion term. The noise $\Gamma (t)$ is assumed to be Gaussian, $\delta$-correlated and does not significantly depend on the state $x(t)$, i.e. we can set $g(x(t), t) = const. = \sigma$ in each window. \\
A change of the sign of the slope 
\begin{align}
\zeta = \left. \frac{\text{d}h(x)}{\text{d}x}\right\vert_{x = x^*} 
\end{align}
of the nonlinear drift at the fixed point $x^*$ which is assumed to be approximately the mean of each window indicates a loss of stability of the fixed point through control parameter change and thus, a bifurcation. We develop $h(x,t)$ into a Taylor series up to order three which is sufficient to describe the normal forms of simple bifurcation scenarios.~\cite{b:strogatz} This results in
\begin{align}
\begin{split}
h(x(t),t) = \alpha_0(t) &+ \alpha_1(t) (x - x^*) + \alpha_2(t) (x - x^*)^2\\
 &+ \alpha_3(t) (x - x^*)^3 + \mathcal{O}(x^4)  \label{eq:taylor} 
\end{split}
\end{align}
so that the information on the linear stability is incorporated in $\alpha_1$. For practical  reasons equation \ref{eq:taylor} in the numerical approach is used in the form
\begin{align}
\begin{split}
    h'_{\rm MC}(x(t),t) = \theta_0 (t; x^*) &+ \theta_1 (t; x^*) \cdot x + \theta_2 (t; x^*) \cdot x^2\\ &+ \theta_3 (t; x^*) \cdot x^3 + \mathcal{O}(x^4),
    \end{split}
\end{align}
where an arbitrary fixed point $x^*$ is incorporated in the coefficients $\underline{\theta}$ by algebraic transformation and comparison of coefficients. 
The estimation of the model parameters $\theta_i$ and $\sigma$ is realized via a MCMC method to reconstruct the full posterior distribution of the drift slope $\zeta$ and the noise level $\sigma$. Starting with Bayes' theorem
\begin{equation}
p(\underline{\theta} |\underline{d}, \mathcal{I}) = \frac{p(\underline{d}|\underline{\theta}, \mathcal{I}) \cdot p(\mathcal{M}|\mathcal{I})}{p(\underline{\theta} | \mathcal{I})}
\end{equation}
via the probability densities $p(.)$, the model parameters $\underline{\theta}$, the time series data $\underline{d}$, the background information $\mathcal{I}$ and the model $\mathcal{M}$, the posterior $p(\underline{\theta} |\underline{d}, \mathcal{I})$ is computed with the product of likelihood and prior $p(\underline{d}|\underline{\theta}, \mathcal{I}) \cdot p(\mathcal{M}|\mathcal{I})$ over the evidence $p(\underline{\theta} | \mathcal{I})$ for normalization.\\ 
The short term propagator
\begin{align}
p(x, t + \tau | x', t) = \frac{1}{\sqrt{2\pi g^2(x',t)\tau }}  \exp\left(-\frac{[x - x' - h(x',t)\tau]^2}{2 g^2(x',t)\tau}\right)
\end{align}
for subsequent times $t$ and $t'$ with $\tau = t - t' \longrightarrow 0$ can be derived from the Langevin equation if the difference $x - x'$ in the exponential expression is approximately defined by the first differences of a given time series and represents the likelihood.
The priors are given by 
\begin{align}
p_{\rm prior}(\theta_0,\theta_1) = \frac{1}{2\pi (1+\theta_1^2)^\frac{3}{2}}
\end{align}
in a broad range of $[-50,50]$ for the linear part of the drift function and Jeffreys' scale prior~\cite{b:linden2014}
\begin{align}
p_{\rm prior}(\sigma ) = \frac{1}{\sigma}
\end{align}
for the noise level $\sigma$ in $[0,50]$. These assumptions reflect best the realistic situation of no or just poor prior information and guarantees the determination of the posterior due to the available data instead of strong prior assumptions. It is taken care that the parameters of the higher orders are initially able to contribute in a similar magnitude to the deterministic dynamics as the linear ones by the priors
\begin{align}
&p_{\rm prior}(\theta_2) = \mathcal{N}(\mu = 0, \sigma = 4), \\ &p_ {\rm prior}(\theta_3) = \mathcal{N}(\mu = 0, \sigma = 8),
\end{align}
with Gaussian distributions $\mathcal{N}$ centred around the mean $\mu = 0$ with a standard deviation $\sigma = 4$ and $\sigma = 8$.\\

The MCMC affine-invariant ensemble sampler of the \textit{emcee} python package is used to compute the posterior. More details about the algorithm and its implementation in the \textit{emcee} package can be found in Foreman-Mackey et al.~\cite{a:Foreman-Mackey2013}. Based on the estimated joint posterior pdf $p(\underline{\theta}|\underline{d},\mathcal{I})$ the parameters $\underline{\theta}$ are sampled and corresponding drift slopes $\zeta$ in the fixed point $x^*$ are calculated.\\
The credibility intervals of the slopes and noise levels are defined as the $\SI{16}{\percent}$ to $\SI{84}{\percent}$- and $\SI{1}{\percent}$ to $\SI{99}{\percent}$-percentiles of the corresponding posterior pdfs. It is computed from a kernel density estimate of the corresponding pdfs. The kernel density estimation is performed with 
$scipy.stats.gaussian\_ kde$~\cite{a:Virtanen2020} using Silverman's rule of thumb to determine the kernel bandwidth.\\
\section{Outlier correction in the post-outage interval}\label{subsec: outlier correction}
The time series $\omega_{\rm R}(t)$ was carefully compared to its printed source in the supplementary material~\cite{a:supplNtip}, section \ref{sec: outlier}, figure \ref{fig: outlier}. The comparison revealed that the outliers were erroneously generated by the image processing tool which misinterpreted grid lines of the scale paper as actual data points. For that reason, we first thinned the original time series by a factor of two, which already cancelled out four of six prominent outliers. Second, we corrected for the remaining two prominent outlier regions by a heuristic approach. The detected outlier intervals, the first and second of which contain four and two points, respectively, were replaced by a sequence of random numbers in a range based on the data observed before and after the intervals. Therefore, we considered the non-outlier samples in a symmetric $\epsilon$-environment of roughly two minutes centered around the outlier peaks. The peaks are subjectively defined by thresholds $T_{\rm out} = 0.01$ and $T_{\rm out} = 0.04$ for the first and second outlier interval, respectively. The data $\omega \in \epsilon$ are used to construct Gaussian kernel density estimates with Silverman's rule. The correction values are drawn from the resulting probability densities. The approach is chosen to avoid making assumptions about the frequency dynamics which we want to analyse. In light of this, the corrections accounts for the magnitude of the outliers, but cannot reconstruct information about the underlying dynamics.

\bibliography{sample}

\section*{Acknowledgments}
M.H. thanks James King (FOIA Public Liaison, BPA), Brian Roth (FOIA Case Coordinator, BPA), Mary Schaff (Librarian, Washington State Library), the anonymous employees of BPA and all the other people who took part in finding the metadata of the two frequency time series for their gentle, cooperative and very dedicated work (cf. supplementary material~\cite{a:supplNtip}, section \ref{sec: metadata}). M.H. thanks the authors from the articles~\cite{a:Hines2011, a:cotilla-sanchez12} for providing the restoration time interval data $\omega_{\rm R}(t)$ used therein and for fruitful discussions. M.H. thanks the Studienstiftung des deutschen Volkes for a scholarship including financial support. O.K. thanks Frank Ehebrecht for co-working to gain the thinned restoration interval dataset $\omega_{\rm R}(t)$ during previous projects (under the old time scale assumptions). We thank colleagues and friends for proofreading the manuscript.

\section*{Author contributions statement and additional information}

M.H. and O.K. have designed research. M.H. performed research, analysed data, designed graphics and modelling thoughts, wrote the manuscript and searched for the frequency time series metadata. O.K. provided thinned restoration interval data $\omega_{\rm R}(t)$ from previous works. The authors declare no competing interests.

\end{document}